\documentclass[aps,a4paper,twocolumn,english, superscriptaddress,eqsecnum,showpacs,longbibliography]{revtex4-1}
\usepackage{babel}
\usepackage{xcolor}
\usepackage{graphicx}
\usepackage{amsmath}
\usepackage{amssymb}
\def\be{\begin{equation}} \def\ee{\end{equation}}
\def\bea{\begin{eqnarray}} \def\eea{\end{eqnarray}}

\definecolor{darkblue}{rgb}{0.1,0.2,0.6}
\definecolor{darkred}{rgb}{0.8,0.1,0.2}
\usepackage[colorlinks,
            citecolor=darkblue,
            linkcolor=darkred,
            urlcolor=darkblue]{hyperref}

\usepackage[all]{hypcap} 

\makeatletter
\adddialect\l@English\l@english
\makeatother

\begin{document}
\title{Dynamical response and dimensional crossover\\
for spatially anisotropic  antiferromagnets}
\author{Maxime Dupont}
\affiliation{Laboratoire de Physique Th\'eorique, IRSAMC, Universit\'e de Toulouse, CNRS, UPS, France}
\author{Sylvain Capponi}
\affiliation{Laboratoire de Physique Th\'eorique, IRSAMC, Universit\'e de Toulouse, CNRS, UPS, France}
\author{Nicolas Laflorencie}
\affiliation{Laboratoire de Physique Th\'eorique, IRSAMC, Universit\'e de Toulouse, CNRS, UPS, France}
\author{Edmond Orignac}
\affiliation{Universit\'e de Lyon, \'Ecole Normale Sup\'erieure de Lyon, Universit\'e Claude Bernard, CNRS, Laboratoire de Physique, F-69342 Lyon, France}

\begin{abstract}
Theoretically challenging, the understanding of the dynamical response in quantum antiferromagnets is of great interest, in particular for both inelastic neutron scattering (INS) and nuclear magnetic resonance (NMR) experiments. In such a context, we theoretically address this question for quasi-one-dimensional quantum magnets, {\it{e.g.}} weakly coupled spin chains for which many compounds are available in Nature. In this class of systems, the dimensional crossover between a three-dimensional ordered regime at low temperature towards one-dimensional physics at higher temperature is a non-trivial issue, notably difficult concerning dynamical properties. Here we present a comprehensive theoretical study based on both analytical calculations (bosonization + random phase and self-consistent harmonic approximations) and numerical simulations (quantum Monte Carlo + stochastic analytic continuation) which allows us to describe the \textit{full temperature crossover} for the NMR relaxation rate $1/T_1$, from one-dimensional Tomonaga-Luttinger liquid physics to the three-dimensional ordered regime, as a function of inter-chain couplings. The dynamical structure factor, directly probing the INS intensity, is also computed in the different regimes.
\end{abstract}

\maketitle

\section{Introduction}\label{sec:introduction}

Among condensed matter systems, the numerous experimental realizations of Mott insulators provide one of the most ideal playgrounds to challenge theoretical descriptions regarding quantum magnetism. For instance, they can realize the widest range of phases of matter, from the most traditional antiferromagnetic (AF) ordering to the most exotic ones such as spin liquids~\cite{balents_spin_2010} or valence bond solids~\cite{affleck_rigorous_1987}. In all cases, dimensionality D plays a crucial role as the effect of quantum fluctuations increases when D is lowered. The best known example of that being Mermin-Wagner-Hohenberg theorem~\cite{mermin1966,hohenberg1967} (and its extensions~\cite{momoi1996,bruno2001}), preventing AF ordering in one dimension, but safely allowing it in two and three dimensions at respectively zero and finite temperature.

While most compounds are intrinsically three dimensional, spatial anisotropies in the energy couplings between degrees of freedom can effectively reduce their effective dimension.
More precisely, what defines the relevant energy scale in a system and therefore its effective dimension is the ratio between temperature and coupling. Considering for example a purely one-dimensional spin systems with  AF exchange coupling $J>0$, while at high temperature $T\gg J$ the system behaves similarly to a classical paramagnet, one expects universal one-dimensional ($1$D) quantum properties at $T\ll J$. In particular, quantum critical chains can be described by the universal Tomonaga-Luttinger liquid (TLL) field theory~\cite{haldane1981,giamarchi2004} in the low-temperature limit, where the physical properties of the system are fully characterized by two parameters, $u$, the  velocity of the excitations, and $K$, a dimensionless parameter. Systems falling into this theoretical description show very peculiar physics and one can ask what is this so-called ``low-temperature limit'' by properly and quantitatively defining the correct low-temperature regime $T\ll J$. This is particularly relevant for realistic quasi-one-dimensional materials where residual couplings are always present, thus inevitably escaping the theoretical one-dimensional world.

For instance,  a 3D array of weakly coupled spin chains with a coupling $J$ along the chains and $J_\perp\ll J$ in the transverse directions is expected to display three dimensional behavior for $T\lesssim J_\perp$,  developing true long-range order. However, at higher temperature this system should exhibit signatures of one-dimensional physics, approximatively in the range $J_\perp\ll T\ll J$. This regime has been already identified for several compounds through thermodynamic quantities. For example the specific heat in the quasi-one-dimensional spin-$1/2$ chain antiferromagnet BaCo$_2$V$_2$O$_8$ material~\cite{seiichiro2008} and in the metal-organic $S=1/2$ two-legs ladder system (C$_5$H$_{12}$N)$_2$CuBr$_4$~\cite{ruegg2008} shows a one-dimensional linear behavior $\propto T$. Another interesting case concerns the (purely 1D) logarithmic corrections predicted by Eggert {\it{et al.}}~\cite{eggert1994} for the magnetic susceptibility of a $S=1/2$ Heisenberg chain, which has been observed for the quasi 1D cuprate Sr$_2$CuO$_3$~\cite{ami1995,eggert1996,motoyama1996}. For weakly coupled two-dimensional planes, how smoothly the ordering process of the three-dimensional system is affected was studied in Ref.~\onlinecite{furuya2016}. This work showed that the AF order parameter $m^\mathrm{AF}(T)$ curve is modified with a non-trivial change of convexity when reducing the interplane coupling as observed in the spin$-1/2$ Heisenberg antiferromagnetic ladder compound (C$_7$H$_{10}$N)$_2$CuBr$_4$ (DIMPY)~\cite{jeong2017}. Similar theoretical works have also been dedicated to the dimensional modulation of the spin stiffness~\cite{laflorencie2012,rancon2017}.

\begin{figure}[t]
    \includegraphics[width=1\columnwidth,clip]{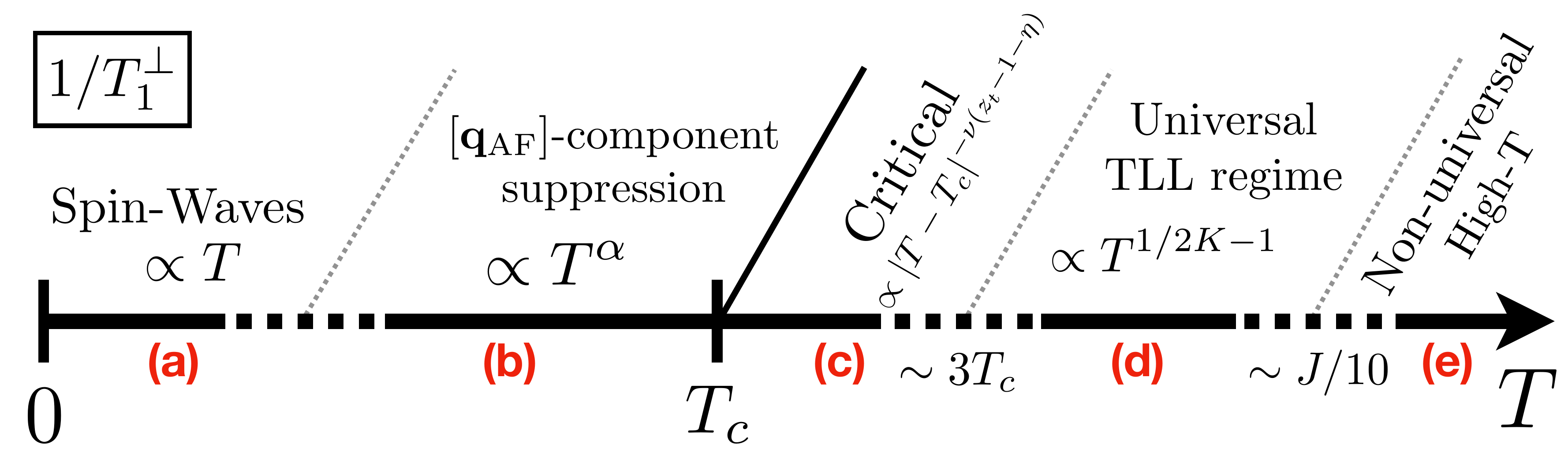}
    \caption{(Color online) Different temperature regimes and crossovers for the transverse component of the NMR relaxation rate $1/T_1^\perp$, as defined in Eq.~\eqref{eq:t1} for an anisotropic three-dimensional antiferromagnet made of weakly coupled chains with an ordering temperature $T_c$. The coupling strengths are $J$ along the chain direction and $J_\perp$ in the transverse direction, see Eq.~\eqref{eq:hamiltonian}. {(a)} Deep in the ordered phase, the NMR relaxation rate increases linearly $\propto T$ from the absolute zero temperature due to spin-waves contributions. {(b)} Right below the critical temperature $T_c$, the NMR relaxation rate goes through a strong algebraic suppression $\propto T^\alpha$ ($\alpha\simeq 4-5$) due to its ``$[\mathbf{q}_\mathrm{AF}]$-component suppression''. The change of behavior from (b) to (a) sets a first crossover temperature. {(c)} When approaching the transition from above the critical temperature, the NMR relaxation rate diverges with critical exponents $\nu$, $\eta$ and $z_t$ characterizing the universality class of the transition, i.e. $\propto |T-T_c|^{-\nu(z_t-1-\eta)}$. The divergence associated to the transition is observed up to approximately $\simeq 3T_c$. {(d)} For $J_\perp/J\ll 1$ we can expect a crossover towards one-dimensional physics with a diverging NMR relaxation rate $\propto T^{1/2K-1}$ where $K$ is the Tomonaga-Luttinger liquid parameter. {(e)} At high temperature, larger than $\sim J/10$, the $1/T_1^\perp$ behavior is non-universal. Note that if $3T_c\gtrsim J/10$, the region (d) of the diagram is squashed, and no universal TLL physics is present in the system, at least regarding the NMR relaxation rate.}
    \label{fig:T1_regimes}
\end{figure}

Nevertheless, it is important to keep in mind that spatial anisotropies only induce dimensional crossover, whereas the true phase transition remains in the same 3D universality class, with a critical ordering temperature $T_c/J\propto \left(J_\perp/J\right)^{\frac{2K}{4K-1}}$~\cite{schulz1996,giamarchi1999}, $K$ being the TLL parameter. Therefore, a key  question we wish to address is about the signatures of a genuine one dimensional physics above $T_c$, and in particular the temperature range where a universal TLL regime is expected. As seen in the ladder system (C$_5$H$_{12}$N)$_2$CuBr$_4$~\cite{ruegg2008} the TLL crossover regime based on measurements of the magnetocaloric effect is not sharply defined. Thus,  one might ask how such a crossover shows up in dynamical quantities such as the dynamical spin structure factor $S_\mathbf{q}(\omega)$ measured by inelastic neutron scattering (INS) experiments, the electron spin resonance (ESR) spectrum~\cite{furuya2015} or the nuclear magnetic resonance (NMR) spin-lattice relaxation rate $1/T_1$. This is  of great experimental interest, in particular to estimate the TLL parameter $K$. For instance, the NMR relaxation rate of a TLL diverges algebraically at low temperature~\cite{klanjsek2008,bouillot2011}
\be
1/T_1\propto T^{1/2K-1}.
\label{eq:T1_1D}
\ee
For a strictly 1D system the crossover temperature separating the non-universal high temperature regime from the low-temperature universal behavior Eq.~\eqref{eq:T1_1D} was recently investigated~\cite{coira2016,dupont2016} using state-of-the-art numerical techniques performing real-time evolution at finite temperature. The authors found that one can indeed asymptotically observe the predicted power-law dependence Eq.~\eqref{eq:T1_1D}, but only at quite low temperature: $T\lesssim J/10$. As for static quantities, a finite three-dimensional coupling $J_\perp$ will ultimately change the dynamical response when approaching $T_c$. When getting close to $T_c$, we will see that the NMR relaxation rate diverges with a power-law $1/T_1\propto |T-T_c|^{-\nu(z_t-1-\eta)}$ with an exponent $\nu(z_t-1-\eta)>0$ characteristic of the phase transition. These different regimes for $T>T_c$ summarized in Fig.~\ref{fig:T1_regimes}\,(c--e) are studied in great detail in this work based on analytical and numerical calculations.

The TLL prediction Eq.~\eqref{eq:T1_1D} is often used to fit the experimentally measured NMR relaxation rate versus $T$ and obtain the dimensionless TLL parameter $K$, but a proper definition of the temperature window inside which the genuine one-dimensional properties can be observed is missing. For instance, in Ref.~\onlinecite{dupont2016} we showed that for the quasi-one-dimensional $S=1$ chain NiCl$_2$-$4$SC(NH$_2$)$_2$ (DTN) material~\cite{mukhopadhyay2012}, the critical temperature is larger than the crossover temperature towards the $1$D regime, thus preventing the observation of TLL behavior. In other words, the region Fig.~\ref{fig:T1_regimes}\,(d) is squashed to zero for DTN, although it has proven to display others $1$D fingerprints~\cite{blinder2017}. Another promising material with a smaller $3$D coupling (hence a smaller $T_c$) is DIMPY~\cite{schmidiger2012,jeong2013} where the $1/T_1$ has been fitted to obtain $K$ versus the external magnetic field $H$, but has shown some discrepancy with the expected value $K(H)$ computed numerically. Our present work reveals that the experimental fitting temperature range $2T_c<T<3T_c$ is probably too close to the critical temperature to be reliable. This will be discussed in greater details in the following.

In NMR experiments, one way to map out the boundary between the disordered and ordered phases is to determine the temperature $T_c$ at which the hyperfine splitting of ``the NMR line'' in the spectrum of the targeted nucleus vanishes~\cite{klanjsek2008,jeong2017,blinder2017}. Another way is to look at the relaxation rate $1/T_1$ as a function of $T$, expected to diverge at the transition, and resulting in practice in a strong enhancement~\cite{klanjsek2015,klanjsek2015prb}. Below $T_c$, experimental observations of the NMR relaxation rate show that it is greatly suppressed with temperature, empirically fitting an algebraic dependence, $1/T_1 \propto T^\alpha$ with $\alpha\simeq 4-5$ as observed in the two-leg spin-$1/2$ ladder Cu$_2$(C$_5$H$_{12}$N$_2$)$_2$Cl$_4$ compound~\cite{mayaffre2000}, DTN~\cite{blinderthesis} and DIMPY~\cite{jeong2017}. This behavior, reported as ``$[\mathbf{q}_\mathrm{AF}]$-component suppression'' in Fig.~\ref{fig:T1_regimes}\,(b) will be discussed and compared with our numerical results, providing some insights and explanations. Finally, although it remains very challenging to observe, both experimentally and numerically since it should happen at very low temperature, deep in the ordered phase the NMR relaxation rate is expected to grow linearly with $T$ due to spin-waves contribution, as shown in Fig.~\ref{fig:T1_regimes}\,(a).

The rest of the paper is organized as follows. In Sec.~\ref{sec:models_and_definitions}, we introduce the theoretical models and provide useful definitions regarding the dynamical quantities of interest in NMR and INS experiments. The numerical techniques as well as the theoretical framework are also briefly described. Section~\ref{sec:results} presents our results for the NMR relaxation rate and the dynamical spin structure factor in weakly coupled spin chains. The different temperature regimes summarized in Fig.~\ref{fig:T1_regimes} are discussed. Finally, we present our conclusions in Sec.~\ref{sec:summary_conclusions}.

\section{Models and definitions}\label{sec:models_and_definitions}

\begin{figure}
    \includegraphics[width=1\columnwidth,clip]{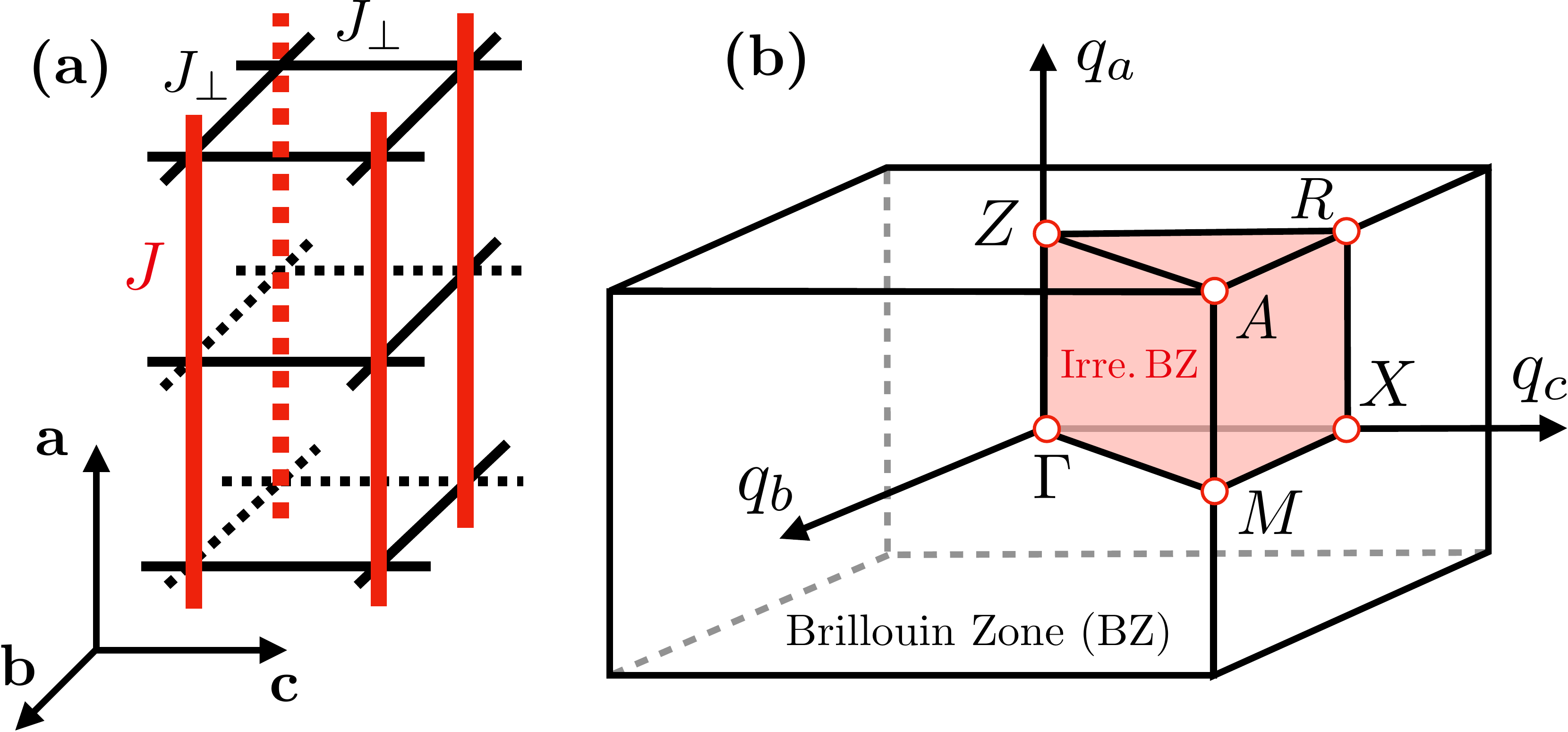}
    \caption{(Color online) (a) Three-dimensional tetragonal lattice with the spatial $\mathbf{a}$ direction non equivalent to $\mathbf{b}$ and $\mathbf{c}$. The spin$-1/2$ degrees of freedom live on the vertices. The Brillouin zone and irreducible Brillouin zone (red region) are shown in panels {(b)} with wave vectors $\mathbf{q}=(q_a, q_b, q_c)$. The vertices $Z=(\pi,0,0)$, $R=(\pi,0,\pi)$, $A\equiv\mathbf{q}_\mathrm{AF}=(\pi,\pi,\pi)$, $\Gamma=(0,0,0)$, $X=(0,0,\pi)$ and $M=(0,\pi,\pi)$ are high-symmetry points of the Brillouin zone.}
    \label{fig:lattice_bz}
\end{figure}

We study coupled quantum spin-$1/2$ chains in three dimensions, ultimately forming a tetragonal lattice as shown in Fig.~\ref{fig:lattice_bz}\,(a). The system is generically described by the following Hamiltonian,
\begin{align}
    \mathcal{H}=\mathcal{H}^\mathrm{1D}+J_\perp\sum_\mathbf{r}\sum_{\mathbf{u}=\mathbf{b},\mathbf{c}}\mathbf{S}_\mathbf{r}\cdot \mathbf{S}_{\mathbf{r}+\mathbf{u}},
    \label{eq:hamiltonian}
\end{align}
where the second term couples nearest-neigbor spins along the transverse directions $\mathbf{b}$ and $\mathbf{c}$ with a Heisenberg interaction of strength $J_\perp$. The first term of Eq.~\eqref{eq:hamiltonian} describes a single XXZ spin chain,
\begin{align}
    \mathcal{H}^\mathrm{1D}=J\sum_\mathbf{r} \left(S^x_\mathbf{r}S^x_{\mathbf{r} + \mathbf{a}} + S^y_\mathbf{r}S^y_{\mathbf{r} + \mathbf{a}} + \Delta S^z_\mathbf{r}S^z_{\mathbf{r} + \mathbf{a}}\right),
    \label{eq:hamiltonian1D}
\end{align}
with $J$ the nearest-neigbor antiferromagnetic exchange along the chain direction $\mathbf{a}$ and $\Delta$ the Ising anisotropy along the $z$ spin component. Although we focus in this work on this specific model, it should apply for any system describing coupled one-dimensional Tomonaga-Luttinger liquids~\cite{giamarchi2004}. The periodicity and spatial symmetries of the system define the Brillouin zone and irreducible Brillouin zone as shown in Fig.~\ref{fig:lattice_bz}\,(b). One can define momentum space spin operators through a Fourier transformation,
\begin{align}
    S^\mu_\mathbf{q}=\frac{1}{\sqrt{N}}\sum_\mathbf{r}\mathrm{e}^{-i\mathbf{q}\cdot\mathbf{r}}S^\mu_\mathbf{r}
    \label{eq:spin_fourier}
\end{align}
with $N$ the total number of spins in the system, $\mathbf{q}=(q_a,q_b,q_c)$ the wave vector with $q_{a,b,c}\in]-\pi,\pi]$ its components along the $\textbf{a}$, $\textbf{b}$, $\textbf{c}$ spatial directions and $\mu\in[x,y,z]$ the spin components respectively.

\subsection{Dynamical quantities}

Overall, we are interested in the dynamical properties of quantum antiferromagnets at finite-temperature described by the Hamiltonian~\eqref{eq:hamiltonian}. The central object is the time-dependent correlation function
\begin{align}
    S^{\mu\upsilon}_\mathbf{q}(t)=\langle S^\mu_{-\mathbf{q}}(t)S^\upsilon_\mathbf{q}(0)\rangle - \langle S^\mu_{-\mathbf{q}}(t)\rangle\langle S^\upsilon_\mathbf{q}(0)\rangle,
    \label{eq:corr_function_t}
\end{align}
where $\langle\rangle$ is the thermal average at inverse temperature $\beta=1/T$, i.e. $\langle\mathcal{O}\rangle=\mathrm{Tr}\,(\mathcal{O}\mathrm{e}^{-\beta\mathcal{H}})/\mathcal{Z}$ with $\mathcal{Z}=\mathrm{Tr}\,\mathrm{e}^{-\beta\mathcal{H}}$ the partition function and $S^\mu_{\mathbf{q}}(t)=\mathrm{e}^{i\mathcal{H}t}S^\mu_{\mathbf{q}}\mathrm{e}^{-i\mathcal{H}t}$ in the Heisenberg representation. Its Fourier transform to frequency space gives the dynamical spin structure factor,
\begin{align}
   S^{\mu\upsilon}_\mathbf{q}(\omega)=\int_{-\infty}^{+\infty}\mathrm{d}t\,\mathrm{e}^{i\omega t} S^{\mu\upsilon}_\mathbf{q}(t),
   \label{eq:corr_function_om}
\end{align}
which is the main quantity of interest throughout this work and is directly related to experimental probes such as INS measurements or the NMR relaxation rate. The static spin structure factor is recovered when integrating over frequencies,
\begin{align}
    S^{\mu\upsilon}_\mathbf{q}=\frac{1}{\pi}\int^{+\infty}_{-\infty}\mathrm{d}\omega\;S^{\mu\upsilon}_\mathbf{q}(\omega),
    \label{eq:static_strfact}
\end{align}
with $\sum_\mathbf{q} S^{\mu\upsilon}_\mathbf{q}=\delta_{\mu\upsilon}/4$ fulfilling the sum rule. It relates to the modulus square of the complex order parameter $m^\mathrm{AF}$,
\begin{align}
    |m^\mathrm{AF}|^2=S^{xx}_{\mathbf{q}_\mathrm{AF}}+S^{yy}_{\mathbf{q}_\mathrm{AF}},
\end{align}
accouting for antiferromagnetic order in the XY plane with $\mathbf{q}_\mathrm{AF}=(\pi,\pi,\pi)$ the antiferromagnetic wave-vector. For convenience and due to the $\mathrm{U}(1)$ symmetry of the Hamiltonian, the transverse part (with respect to the Ising anisotropy direction) can be isolated and written using raising and lowering operators,
\begin{align}
    S^{xx}_\mathbf{q}+S^{yy}_\mathbf{q}=\frac{1}{2}\Bigl(S^{+-}_\mathbf{q}+S^{-+}_\mathbf{q}\Bigr).
\end{align}

\subsubsection{Inelastic Neutron Scattering}

Inelastic neutron scattering is a spectroscopy technique that can directly probe the spectral function Eq.~\eqref{eq:corr_function_om}. The wave vector $\mathbf{q}$ is the momentum transferred to the sample between incoming and outgoing wave vectors $\mathbf{k}'$ and $\mathbf{k}$ of the neutrons, and $\omega$ the kinetic energy transferred to the system due to the collision. More precisely, INS experiments measure the partial differential cross section~\cite{furrer2009,zaliznyak2015},
\begin{align}
    \frac{\mathrm{d}^2\sigma(\mathbf{q},\omega)}{\mathrm{d}\Omega\mathrm{d}\omega}=\frac{\|\mathbf{k}'\|}{\|\mathbf{k}\|}F^2_\mathbf{q}\sum_{\mu\upsilon}\left(\delta_{\mu\upsilon}-\frac{q_\mu q_\upsilon}{\mathbf{q}^2}\right)S^{\mu\upsilon}_\mathbf{q}(\omega),
\end{align}
with $q_{\mu}$ the projection of the wave vector $\mathbf{q}$ on the spin component $\mu$. The prefactor of the dynamical spin structure factor in the sum ensures that only the spin components normal to $\mathbf{q}$ contribute to the cross section. The magnetic form factor $F_\mathbf{q}$ is the Fourier transform of the spatial density of the scatterer, i.e. the electrons holding the relevant spin degrees of freedom in our case. The ratio $\|\mathbf{k}'\|/\|\mathbf{k}\|$ and the form factor are known quantities that can be factored out of the experimental data. Furthermore, the spectral function is only non zero if $\mu=\upsilon$ for the Hamiltonian~\eqref{eq:hamiltonian}, which gives the corrected scattering intensity,
\begin{align}
    I=\sum_\mu=\left(1-\frac{q_\mu^2}{\mathbf{q}^2}\right)S^{\mu\mu}_\mathbf{q}(\omega)=I^\perp+I^\parallel,
\end{align}
where the intensity has been separated into longitudinal and transverse parts due to the $\mathrm{U}(1)$ symmetry of the Hamiltonian. Moreover, the $q_\mu$-dependent prefactors can be calibrated in experimental setups and will be set to unity in the following, which results in
\begin{align}
    I^\parallel=S^{zz}_\mathbf{q}(\omega),~\mathrm{and}~I^\perp=\frac{1}{2}\Bigl[S^{+-}_\mathbf{q}(\omega)+S^{-+}_\mathbf{q}(\omega)\Bigr].
    \label{eq:ins_intensity}
\end{align}
We shall focus on the transverse contribution in the following, related to the antiferromagnetic XY order below the critical temperature $T_c$.

\subsubsection{NMR spin-lattice relaxation rate}

In NMR experiments, the nuclear spins of the sample are polarized through an external magnetic field and then perturbed by an electromagnetic pulse. One can select and target specific nuclei by choosing the right frequency $\omega_0$ corresponding to the level splitting of the picked nuclei due to Zeeman effect. Following the perturbation, the nuclear spins precess around the magnetic field direction and relax over time with an energy transfer to the external environment, the lattice and specifically the electrons~\cite{abragam1961,horvatic2002,slichter2013}. The return of magnetization $M$ to equilibrium reads $1-M(t)\propto\mathrm{e}^{-t/T_1}$, where $1/T_1$ is known as the spin-lattice relaxation rate and can be related to the dynamical correlation function in crystalline magnets,
\begin{align}
    \frac{1}{T_1}=\sum_\mathbf{q}\sum_{\mu\upsilon}(A^{\mu\upsilon}_\mathbf{q})^2S^{\mu\upsilon}_\mathbf{q}(\omega_0),
    \label{eq:t1_def}
\end{align}
with $A^{\mu\upsilon}_\mathbf{q}$ the hyperfine tensor describing the dipolar interaction between nuclear and electronic spins. Its $\mathbf{q}$-dependence provides a kind of form factor which can modify the sensitivity of $1/T_1$ to different wave vector components of the spin dynamics~\cite{thurber2001,sirker2009,sirker2011}, although for generality, we will consider it to be independent of $\mathbf{q}$ and equal to unity in the following. Thereby, the sum over $\mathbf{q}$ simplifies to the local ($\mathbf{r}=0$) dynamical correlation function in real space,
\begin{align}
    \frac{1}{T_1}=\sum_{\mu\upsilon}S^{\mu\upsilon}_{\mathbf{r}=\mathbf{0}}(\omega_0)=\frac{1}{T_1^\perp}+\frac{1}{T_1^\parallel},
    \label{eq:t1_def2}
\end{align}
and is only non zero for $\mu\equiv\upsilon$ for the Hamiltonian~\eqref{eq:hamiltonian}. As for the scattering intensity, we can separate the longitudinal and transverse contributions,
\begin{align}
    \frac{1}{T_1^\parallel}=S^{zz}_{\mathbf{r}=\mathbf{0}}(\omega_0),~\mathrm{and}~\frac{1}{T_1^\perp}=\frac{1}{2}\Bigl[S^{+-}_{\mathbf{r}=\mathbf{0}}(\omega_0)+S^{-+}_{\mathbf{r}=\mathbf{0}}(\omega_0)\Bigr].
    \label{eq:t1}
\end{align}
It is theoretically justified to take the limit $\omega_0\rightarrow 0$ since the NMR frequency is of a few tens or hundreds of MHz, corresponding to temperatures of the order of mK, often making it the smallest energy scale of the problem. However, taking this limit supposes some smoothness in the local spectral function $S^{\mu\upsilon}_{\mathbf{r}=\mathbf{0}}(\omega)$, with no sharp contribution at $\omega\rightarrow 0$ that would not be captured by the actual NMR measurements due to the finiteness of $\omega_0$. As for the scattering intensity, we will focus on the transverse contribution which dominates over the longitudinal one from intermediate to low temperatures.

\subsection{Numerical methods}

The dynamical correlation functions $S^{\mu\upsilon}_\mathbf{q}(\omega)$ are first computed in imaginary time following Ref.~\onlinecite{dorneich2001}, using the stochastic series expansion quantum Monte Carlo (QMC) method~\cite{sandvik1991,sandvik2010} with operator-loop update~\cite{sandvik1999}. The analytic continuation from imaginary-time to real-frequency correlation functions is then performed using the Stochastic Analytic Continuation (SAC) method~\cite{sandvik1998,syljuasen2008,fuchs2010,wu2013}. In all cases, the general idea is to invert the following equation
\begin{align}
    S_\mathbf{q}^{\mu\upsilon}(\tau)=\frac{1}{\pi}\int^{+\infty}_{-\infty}\mathrm{d}\omega\;\mathrm{e}^{-\tau\omega}\;S_\mathbf{q}^{\mu\upsilon}(\omega),
    \label{eq:sac_def}
\end{align}
where $\tau$ ($\equiv-it$) is the imaginary time. The main difficulty into inverting this relationship lies in the fact that only a QMC estimate of $S^{\mu\upsilon}_\mathbf{q}(\tau)$ is available with its intrinsic statistical sampling error. Thereby, in practice, only broad features of $S_\mathbf{q}^{\mu\upsilon}(\omega)$ can be resolved since the information on its fine structure is only present at a level of precision in the imaginary-time correlation functions that is not achievable in numerical simulations. In an attempt to overcome this issue, the basic idea of the stochastic analytic continuation is to represent the spectrum by a large number of delta peaks which positions are sampled in order to provide a good fit (in a chi-square sense) of the imaginary-time data. More precisely, starting with an initial representation of $S_\mathbf{q}^{\mu\upsilon}(\omega)$, the $\delta$-functions are moved by means of a standard Metropolis algorithm with a probability of acceptance
\begin{align}
    \mathcal{P}\Bigl[S_\mathbf{q}^{\mu\upsilon}(\omega)\Bigr]\propto\mathrm{exp}\Bigl(-\chi^2/2\Theta\Bigr),
\end{align}
where $\chi^2$ measures the goodness of the fit between the imaginary time QMC data and the one obtained from $S_\mathbf{q}^{\mu\upsilon}(\omega)$ after a $\delta$-peak move using expression~\eqref{eq:sac_def}. The parameter $\Theta$ is the sampling temperature, optimized using Bayesian inference~\cite{fuchs2010}. Overall, more technical details are available in Refs.~\onlinecite{sandvik1998},~\onlinecite{syljuasen2008} and~\onlinecite{shao2017}.

\subsection{Quasi-one-dimensional physics}

\subsubsection{Bosonization of independent chains}\label{sec:bosonization}

The low energy, long wavelength physics of the one-dimensional Hamiltonian~\eqref{eq:hamiltonian1D} for an Ising anisotropy $|\Delta|<1$ can be captured by the bosonization formalism~\cite{luther1975,haldane1980,dennijs1981}. Setting the lattice spacing to unity, the bosonized Hamiltonian reads
\begin{align}
    \label{eq:bosonized1D}
    \mathcal{H}=\sum_{\mathbf{r}_\perp}\int \frac{\mathrm{d}x}{2\pi}\Biggl\{uK\Bigl[\pi \Pi_{\mathbf{r}_\perp}(x)\Bigr]^2 + \frac{u}{K}\Bigl[\partial_x \phi_{\mathbf{r}_\perp}(x)\Bigr]^2\Biggr\},
\end{align}
where $\mathbf{r}_\perp=n_b \mathbf{b} + n_c \mathbf{c}$, $x$ is the position along the direction $\mathbf{a}$, and $[\phi_{\mathbf{r}_\perp}(x),\Pi_{\mathbf{r'}_\perp}(x')]=i \delta_{\mathbf{r}_\perp,\mathbf{r'}_\perp} \delta(x-x')$. The velocity $u$ and the Tomonaga-Luttinger exponent $K$ are related to the microscopic model parameters $J$ and $\Delta$ by~\cite{luther1975},
\begin{align}
    K=\frac{\pi}{2\arccos(-\Delta)},~\mathrm{and}~
    u=\frac{J\pi\sqrt{1-\Delta^2}}{2\arccos \Delta }.
    \label{eq:tll_parms}
\end{align}
The spin operators can be represented in terms of the fields of the bosonized Hamiltonian~\cite{luther1975,haldane1980} and yields in the critical TLL regime $-1<\Delta\le 1$ to quasi-long-range order for spin-spin correlations, i.e. they decay as a power law with the distance $d$ at zero temperature~\cite{luther1975,haldane1980,dennijs1981},
\begin{eqnarray}
    \label{eq:chain-t0-pm}
    \langle S^\pm_x S^\mp_{x+d}\rangle &=&(-1)^{d}A_\perp d^{-\frac 1 {2K}}-\tilde{A}_\perp d^{-2K-\frac 1 {2K}}, \\
    \label{eq:chain-t0-zz}
    \langle S^z_x S^z_{x+d} \rangle &=&  -\frac K {2\pi^2} d^{-2} + (-1)^d \frac{A_{\parallel}}{2} d^{-2K},
\end{eqnarray}
where the parameters $A_\perp$, $\tilde{A}_\perp$ and $A_\parallel$ can be expressed as a function $K$~\cite{lukyanov1999,lukyanov2003,hikihara2004} (see appendix~\ref{app:amplitudes}). At finite temperature, the decay becomes exponential with a correlation length $\sim u/T$ and dynamical spin structure factors Eq.~\eqref{eq:corr_function_om} have been obtained~\cite{schulz1983,schulz1986,giamarchi2004} in terms of Euler Beta
functions~\cite{abramowitz1972}. NMR
relaxation rates~\eqref{eq:t1_def} have been found~\cite{sachdev1994,chitra1997,barzykin2001,klanjsek2008} in the form
\begin{eqnarray}
\label{eq:ll_spsm_corr}
    \frac{1}{T_1^\perp} &=&
     \frac{A_\perp\cos\left(\frac{\pi}{4K}\right)B\left(\frac{1}{4K},1-\frac{1}{2K}\right)}{u}\left(\frac{2\pi T}{u}\right)^{\frac{1}{2K}-1} \\
    \frac{1}{T_1^\parallel} &=& \frac{A_\parallel\cos\left(\pi K\right)B\left(K,1-2K\right)}{2u}\left(\frac{2\pi T}{u}\right)^{2K-1}\nonumber\\
    &+&\frac{KT}{4\pi u^2},
    \label{eq:ll_szsz_corr}
 \end{eqnarray}
with $B(x,y)$ the Euler beta function and $A_{\perp,\parallel}$ prefactors of the static correlation functions appearing in Eqs.~\eqref{eq:chain-t0-pm} and \eqref{eq:chain-t0-zz}.  In the 1D critical regime  $K\ge 1/2$  $1/T_1^\perp(T)$ diverges at zero temperature as a $K$-dependent power-law, and dominates over $1/T_1^\parallel$ which vanishes at low-temperature. The analytical prediction for $1/T_1^{\perp}$ Eq.~\eqref{eq:ll_spsm_corr} has been perfectly checked against numerics~\cite{coira2016,dupont2016}, without any adjustable parameter. The agreement becomes excellent for $T< J/10$.

In the presence of a weak interchain coupling as in Eq.~\eqref{eq:hamiltonian}, the quasi-long range order of the chains will turn into a true long range order for sufficiently low temperature. For $\Delta<1$ ordering in the XY plane is favored. To describe such long-range ordering within bosonization, either the Random Phase Approximation (RPA)~\cite{schulz1996,giamarchi1999} or the self-consistent Approximation (SCHA)~\cite{suzumura1979,donohue2001,cazalilla2006,you2012} can be used. As we will see below, RPA is more convenient to address the fluctuations above the transition, while SCHA gives a simpler picture of the low temperature phase.

\subsubsection{Random Phase Approximation}
\label{sec:rpa}

In the random phase approximation~\cite{schulz1996,giamarchi1999}, one assumes that a spin chain responds to an effective field $h_\mathbf{q}^\mu(\omega)$ that is the sum of the applied space-time dependent external field $h_\mathbf{q}^\mu(\omega)$ and an internal field generated by the sum of responses of the other chains,
\begin{multline}
    \label{eq:rpa-def}
    h_\mathbf{q}^\mu(\omega) =h_\mathbf{q}^\mu(\omega) -2 J_\perp\Bigl[\cos(\mathbf{q}\cdot\mathbf{b})\\
    +\cos(\mathbf{q}\cdot\mathbf{c})\Bigr]\langle S_\mathbf{q}^\mu(\omega)\rangle
\end{multline}
In Eq.~\eqref{eq:rpa-def}, the angle brackets indicate an expectation value calculated in linear response for a single chain immersed in the self-consistent field. Above the critical temperature, the linear response of the single chain is given by $\langle S_\mathbf{q}^\mu(\omega)\rangle = \chi^{\mu\mu}_\mathrm{1D}(q_a,\omega)  h_\mathbf{q}^\mu(\omega)$ with $\chi^{\mu\mu}_\mathrm{1D}(q_a,\omega)$ the susceptibility of a single chain along the $\mathbf{a}$ direction with $\mu\in[x,y,z]$ the spin components. This yields to the following expression for the susceptibility of the three-dimensional system~\cite{schulz1996,giamarchi1999}
\begin{align}
    \label{eq:rpa-disordered}
    \chi^{\mu\mu}_{\mathbf{q}}(\omega)=\frac{\chi^{\mu\mu}_\mathrm{1D}(q_a,\omega)}{1+2J_\perp  [\cos(\mathbf{q}\cdot\mathbf{b})+\cos(\mathbf{q}\cdot\mathbf{c})]
    \chi^{\mu\mu}_\mathrm{1D}(q_a,\omega)},
\end{align}
from which the dynamical spin structure factor $S_{\mathbf{q}}^{\mu\mu}(\omega)=\coth(\beta\omega/2) \mathrm{Im}\chi^{\mu\mu}_{\mathbf{q}}(\omega)$ is obtained using the fluctuation-dissipation theorem. The static response function, $\chi^{\mu\mu}_{\mathbf{q}_\mathrm{AF}}(\omega=0)$ is divergent at a temperature such that $1-4J_\perp \chi^{\mu\mu}_\mathrm{1D}(q_a=\pi,\omega=0)=0$. It can be shown that the divergence occurs at a higher temperature $T_c$ for $\mu=x,y$ than $\mu=z$~\cite{giamarchi1999}. In three dimensions, below that temperature $T_c$, easy-plane antiferromagnetic order sets in, and Eq.~\eqref{eq:rpa-disordered} is not anymore applicable.

In the ordered phase, each chain is in a mean-field $h^x_{\mathbf{r}}=h^x_\mathrm{MF}e^{i \mathbf{q}_\mathrm{AF} \cdot\mathbf{r}}$. As a result, rotation symmetry is reduced to a $\Bbb{Z}_2$ rotation around the $x$ axis, and translation symmetry to even multiples of $\mathbf{a}$. Besides the normal response functions, $\chi_{\mathrm{1D},n}^{\mu\mu}$, an umklapp response $\chi^{yz}_{\mathrm{1D},u}$ is present. The expressions of RPA susceptibility are modified~\cite{schulz1996}, and poles associated with Goldstone modes appear. Such modes are expected to yield a contribution linear in temperature to the NMR relaxation rate $T_1$. However, in order to do precise calculations of response functions within bosonization, since the bosonized Hamiltonian in the ordered phase is a quantum sine-Gordon model~\cite{giamarchi2004,klanjsek2008}, one has to resort to form factor expansion techniques~\cite{karowski1978} generalized to positive temperature~\cite{essler2009}. Such calculations quickly become very involved, and a more elementary approach is provided by the self-consistent harmonic approximation~\cite{suzumura1979,donohue2001,cazalilla2006,you2012}.

\subsubsection{Self-consistent harmonic approximation}
\label{sec:scha}

In the low temperature phase, we have to consider the full Hamiltonian,
\begin{multline}
    \label{eq:bosonized-interchain}
    \mathcal{H} = \mathcal{H}^\mathrm{1D} + \sum_{\mathbf{r}_\perp} J_\perp A_\perp\int \mathrm{d}x \Bigr[\cos(\theta_{\mathbf{r}_\perp}-\theta_{\mathbf{r}_\perp+\mathbf{b}})\\
    + \cos(\theta_{\mathbf{r}_\perp}-\theta_{\mathbf{r}_\perp+\mathbf{c}})\Bigl].
\end{multline}
In the self-consistent harmonic approximation (SCHA)~\cite{suzumura1979}, one
makes the approximation:
\begin{multline}
    \label{eq:scha-approx}
    \cos(\theta_{\mathbf{r}_\perp} -\theta_{\mathbf{r}_\perp+\mathbf{b}})\simeq\Bigl\langle \cos(\theta_{\mathbf{r}_\perp}-\theta_{\mathbf{r}_\perp+\mathbf{b}})\Bigr\rangle\times\\ \Biggl[1-
    \frac{1}{2} (\theta_{\mathbf{r}_\perp}-\theta_{\mathbf{r}_\perp+\mathbf{b}})^2+ \frac{1}{2}\Bigl\langle(\theta_{\mathbf{r}_\perp}-\theta_{\mathbf{r}_\perp +\mathbf{b}})^2\Bigr\rangle\Biggr],
\end{multline}
turning~\eqref{eq:bosonized-interchain} into a quadratic Hamiltonian whose interchain coupling $J_\perp A_\perp \langle\cos(\theta_\mathbf{r'_\perp}-\cos\theta_\mathbf{r_\perp})\rangle$ have to be determined self-consistently \cite{donohue2001,cazalilla2006,you2012}.
The SCHA allows to calculate the expectation value of the order parameter~\cite{donohue2001} and predicts dispersion of Goldstone modes~\cite{cazalilla2006}.

\section{Results}\label{sec:results}
Below we discuss our results for the dynamical structure factor and the NMR relaxation rate, both numerically obtained using large scale QMC+SAC. We divide the discussion in three parts corresponding to the different temperature regimes. Numerical results are compared to analytical predictions when they are available.

\subsection{One-dimensional Tomonaga-Luttinger regime at high temperature}

\subsubsection{RPA expression for the NMR relaxation rate}
According to bosonization~\cite{giamarchi2004}, the NMR relaxation rates~\eqref{eq:t1_def2} can be split into a dominant $q_a \simeq\pi$ and a subdominant $q_a \simeq 0$ contribution.
Focusing on the dominant transverse response at $q_a=\pi$ one gets
\begin{widetext}
\begin{eqnarray}
 \label{eq:t1-qpia-perp}
 \left(\frac 1 {T^\perp_1}\right)_{q_a=\pi,\mathrm{RPA}} &=&   \int \frac{\mathrm{d}^2
   \mathbf{q}_\perp}{(2\pi)^2} \int_{-\Lambda}^\Lambda
   \frac{\mathrm{d}q_a}{2 \pi}
    \frac{\lim_{\omega
   \to 0} \frac T \omega \mathrm{Im} \chi^{\pm}_\mathrm{1D}(\pi + q_a,\omega)}{\left[1+2J_\perp[\cos(\mathbf{q}_\perp\cdot \mathbf{b})+\cos(\mathbf{q}_\perp\cdot\mathbf{c})]
      \mathrm{Re} \chi^{\pm}_\mathrm{1D}(\pi+ q_a,0)\right]^2 }\\
  \label{eq:t1-qpia-tn}
&=&
  \frac{A_\perp}{2u \Gamma^2\left(\frac 1 {4K}\right)} \left(\frac{2\pi T}{u}
  \right)^{\frac 1 {2K}-1} \int_{-\infty}^{+\infty} \frac{\mathrm{d}\xi}{\sin^2\left(\frac {\pi
      } {8K} \right)+ \sinh^2(\pi \xi)}\left|\frac{ \Gamma\left( \frac 1 {8K} + i
          \xi\right)}{\Gamma\left(1-\frac
          1 {8K} + i \xi\right)} \right|^2 \nonumber \\ &&
\qquad\qquad\qquad\qquad\qquad\qquad\qquad\qquad\times\frac{\mathrm{E}\left[ \displaystyle\left(\frac{T_c}{T}\right)^{4-1/K}\left|\frac{\Gamma\left(1-\frac
          1 {8K}\right) \Gamma\left(\frac 1 {8K} + i
          \xi\right)}{\Gamma\left(\frac 1 {8K}\right) \Gamma\left(1-\frac
          1 {8K} + i \xi\right)} \right|^4 \right]} {1- \displaystyle\left(\frac{T_c}{T}\right)^{4-1/K}\left|\frac{\Gamma\left(1-\frac
          1 {8K}\right) \Gamma\left(\frac 1 {8K} + i
          \xi\right)}{\Gamma\left(\frac 1 {8K} \right) \Gamma\left(1-\frac
          1 {8K} + i \xi\right)} \right|^4},
\end{eqnarray}
\end{widetext}
where the integration over $\mathbf{q}_\perp$ has been performed exactly in terms of elliptic integrals~\cite{abramowitz1972}. In Eq.~\eqref{eq:t1-qpia-tn} $\mathrm{E}(x)$ is a complete elliptic integral of the second kind~\cite{abramowitz1972} and $\Gamma(x)$ is the Euler Gamma function. The expression~\eqref{eq:t1-qpia-tn} can be rewritten as:
\begin{equation}
    \label{eq:t1-qpia-scal}
    \left(\frac{1}{T_1}\right)_{q_a=\pi,\mathrm{RPA}} =
    \left(\frac{1}{T_1^\perp}\right)_\mathrm{1D} \times \Phi\left(\frac{T_c}{T},K\right).
\end{equation}
The enhancement factor $\Phi$ in Eq.~\eqref{eq:t1-qpia-scal} depends only on $T_c/T$ and the Tomonaga-Luttinger exponent $K$. In the limit $T_c/T\rightarrow 0$, $\Phi(T_c/T,K)\to 1$, and the single chain behavior Eq.~\eqref{eq:ll_spsm_corr} is recovered.
As we will discuss below, the above RPA expression for
$1/T_1^\perp$ which describes the disordered regime $T>T_c$ can be directly compared with QMC results.

\begin{figure*}[!ht]
    \includegraphics[width=2\columnwidth,clip]{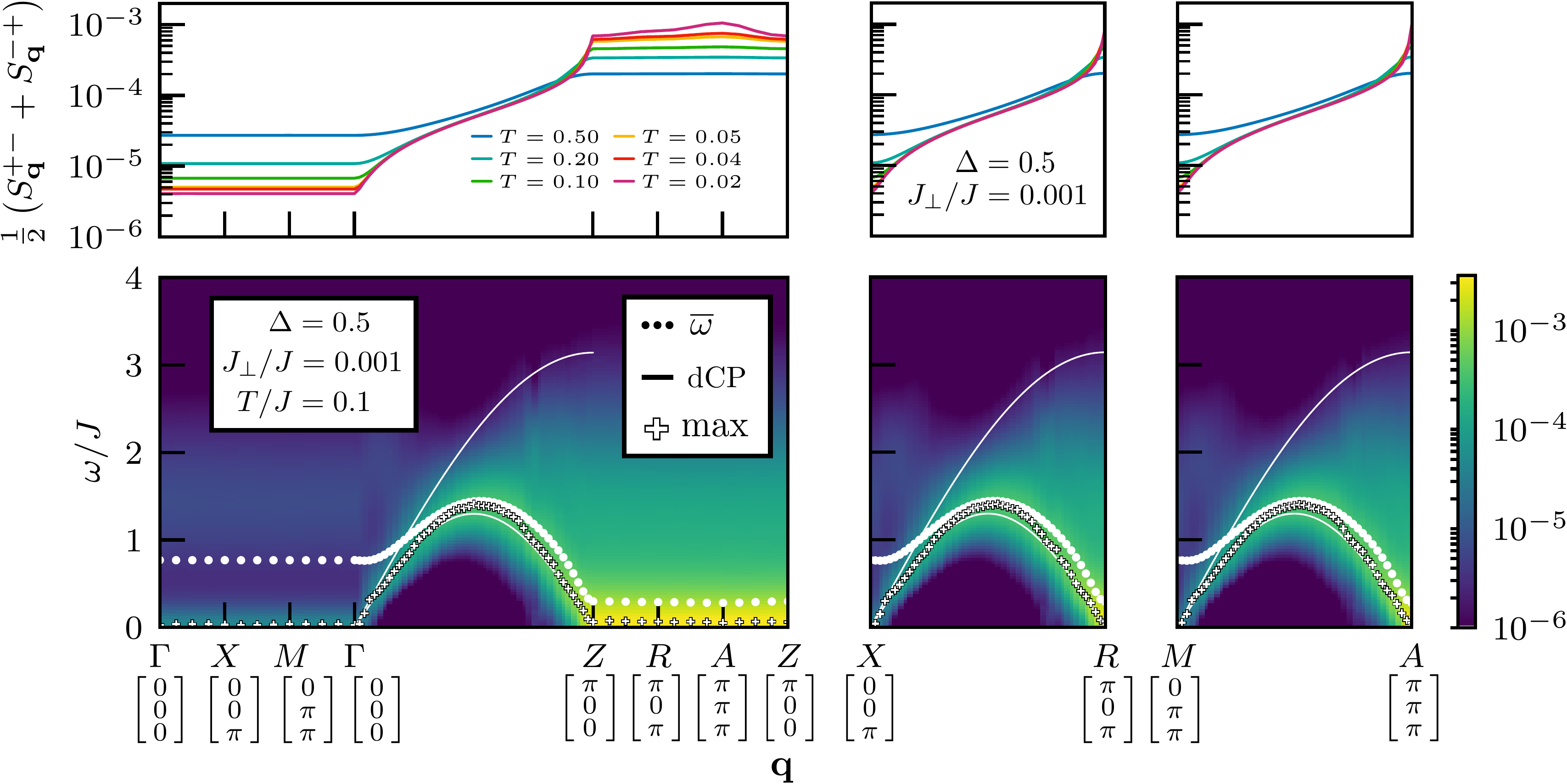}
    \caption{(Color online) The lower panels show the transverse inelastic neutron scattering intensity $I^\perp$ for weakly coupled spin chains in three dimensions with $J_\perp/J=0.001$ and an Ising anisotropy $\Delta=0.5$ along the spatial $\textbf{a}$ direction. The $\mathbf{q}$ points follow the high symmetry lines of the BZ of Fig.~\ref{fig:lattice_bz}\,(b). The temperature of the system is $T=0.1J$, such that we are in the universal one-dimensional regime with $3T_c\lesssim T\lesssim 0.1J$, making the BZ lines $\Gamma Z$, $XR$ and $MA$ equivalent and corresponding all to the single chain spectrum. The white dot symbols show the first moment of the spectrum and the plus symbols the position of the maximum of intensity at a given $\mathbf{q}$ point. We also show the two sine branches of the des Cloizeaux-Pearson dispersion relations in Eq.~\eqref{eq:descloizeauxpearson} where the prefactor of the lower one corresponds to the TLL velocity $u\simeq 1.299J$ of a single chain with $\Delta=0.5$. Note that the critical temperature for this system is $T_c/J\simeq 0.007$. The upper panels correspond to the static structure factor. The data are from quantum Monte Carlo simulations on the largest available system of size $N=96\times 8\times 8=6\,144$ spins.}
    \label{fig:colormap_3d_above_tc}
\end{figure*}

\subsubsection{QMC results}
\paragraph{Dynamical structure factor.---}
Before addressing the NMR relaxation rate, let us first discuss the dynamical structure factor in the paramagnetic regime above the transition. To do so we have simulated very weakly coupled XXZ chains $J_\perp/J=10^{-3}$ in Eq.~\eqref{eq:hamiltonian} and $\Delta=0.5$ in Eq.~\eqref{eq:hamiltonian1D}. Such a very anisotropic system orders below $T_c\simeq 0.007 J$~\footnote{Throughout this work, we basically consider three different three-dimensional coupling strengths. The first one is $J_\perp/J=0.1$ for which the critical temperature is evaluated by computing the spin stiffness~\cite{pollock1987,sandvik1997} using quantum Monte Carlo. It reveals a finite temperature transition using a standard finite-size scaling analysis~\cite{sandvik2010}, $\rho_s(L)=L^{2-D}\,\, \mathcal{G}_{\rho_s}\left[L^{1/\nu}\left(T-T_c\right)\right]$, where $D=3$ is the dimensionality and $\nu=0.6717$ the $3$D-XY critical correlation length exponent~\cite{burovski2006,campostrini2006,beach2005}. We find that in this case, $T_c/J=0.22406(3)$. For the two other cases considered, $J_\perp/J=0.01$ and $J_\perp/J=0.001$, the critical temperature is too small to be evaluated using the previous method. Instead, we compute it using the RPA estimate, based on the susceptibility expression of a single chain in a mean-field three-dimensional environment, i.e. $T_c=\frac{u}{2\pi}\left[\frac{4A_\perp J_\perp\sin\left(\frac{\pi}{4K}\right)B^2(\frac{1}{8K}, 1-\frac{1}{4K})}{2u}\right]^{\frac{2K}{4K-1}}$~\cite{giamarchi2004,chitra1997,bouillot2011}. We explicitly include a renormalization parameter $\alpha$ to take into account the effects of spin fluctuations beyond the mean-field treatment of interchain interaction, i.e. $J_\perp\rightarrow \alpha J_\perp$. This was first discussed analytically for the Heisenberg  spin chain in zero field in Ref.~\onlinecite{irkhin2000} and then precisely verified numerically in Ref.~\onlinecite{yasuda2005}, where $\alpha=0.695$ was obtained. A slightly different value, $\alpha=0.74$ was successfully applied in describing $T_c(H)$ of the (C$_5$H$_{12}$N)$_2$CuBr$_4$ compound~\cite{thielemann2009,bouillot2011}, while for DTN, $\alpha=0.67(2)$ was found~\cite{blinder2017}, pointing to a quite universal value of this correction. Here, we have used the rescaling $J_\perp\rightarrow 0.69J_\perp$, which gives $T_c/J\simeq 0.039$ and $T_c/J\simeq 0.007$ for $J_\perp/J=0.01$ and $J_\perp/J=0.001$ respectively.}. In Fig.~\ref{fig:colormap_3d_above_tc} we show the transverse scattering intensity along the high symmetry lines of the BZ, computed in quantum Monte Carlo supplemented by stochastic analytic continuation. The spectrum along the chains direction $\mathbf{a}$, corresponding to the lines $\Gamma Z$, $XR$ and $MA$ are indistinguishable compared to the single chain spectrum. This is expected for such weakly coupled chains in a temperature range fulfilling $T_c\ll T\ll J$. For comparison, the better-known isotropic $\mathrm{SU}(2)$ Heisenberg chain, where Bethe ansatz calculations are available at zero temperature, has its dominant contribution (i.e. $98\%$ of the spectral weight) coming from a two and four-spinon continuum~\cite{bougourzi1996,karbach1997,caux2005,caux2006} bounded from below and above by des Cloizeaux-Pearson (dCP) dispersion relations~\cite{descloizeaux1962,yamada1969}
\begin{equation}
    \label{eq:descloizeauxpearson}
    \omega_\mathrm{lower}(q) = \frac{J\pi}{2}|\sin q|,\quad    \omega_\mathrm{upper}(q) = J\pi\left|\sin \frac{q}{2}\right|.
\end{equation}
For the XXZ case, predictions are only available for the longitudinal dynamical spin structure factor at small $q$~\cite{pereira2006,pereira2007}, e.g. $S^{zz}_{q\rightarrow 0}=Kq$ with $K$ the dimensionless TLL parameter. Similarities are nonetheless visible: excitations are bounded from above by $\omega_\mathrm{upper}(q)$ and from below by a sine branch with a prefactor corresponding to the TLL velocity $u\simeq 1.299J$ (computed from Eq.~\eqref{eq:tll_parms} for $\Delta=0.5$), a bit smaller than the velocity at the isotropic point, $u=J\pi/2$. The bounds are broadened due to finite temperature effects. Low-energy ($\omega\rightarrow 0$) excitations are restricted here to the usual commensurate modes $q\sim 0$ and $q\sim\pi$, while it is known that in presence of any additional magnetic field along the same direction as the Ising anisotropy (hence at finite magnetization density $m^z$), the XY correlations of the system would develop incommensurate modes at $q=2\pi m^z$ and $q=2\pi(1-m^z)$ in addition to the commensurate ones~\cite{giamarchi2004,chitra1997}. As the temperature is decreased towards $T_c$, the spectral weight (data not shown) gets concentrated more and more around the AF wavevector.

\begin{figure}[!h]
    \includegraphics[width=1\columnwidth,clip]{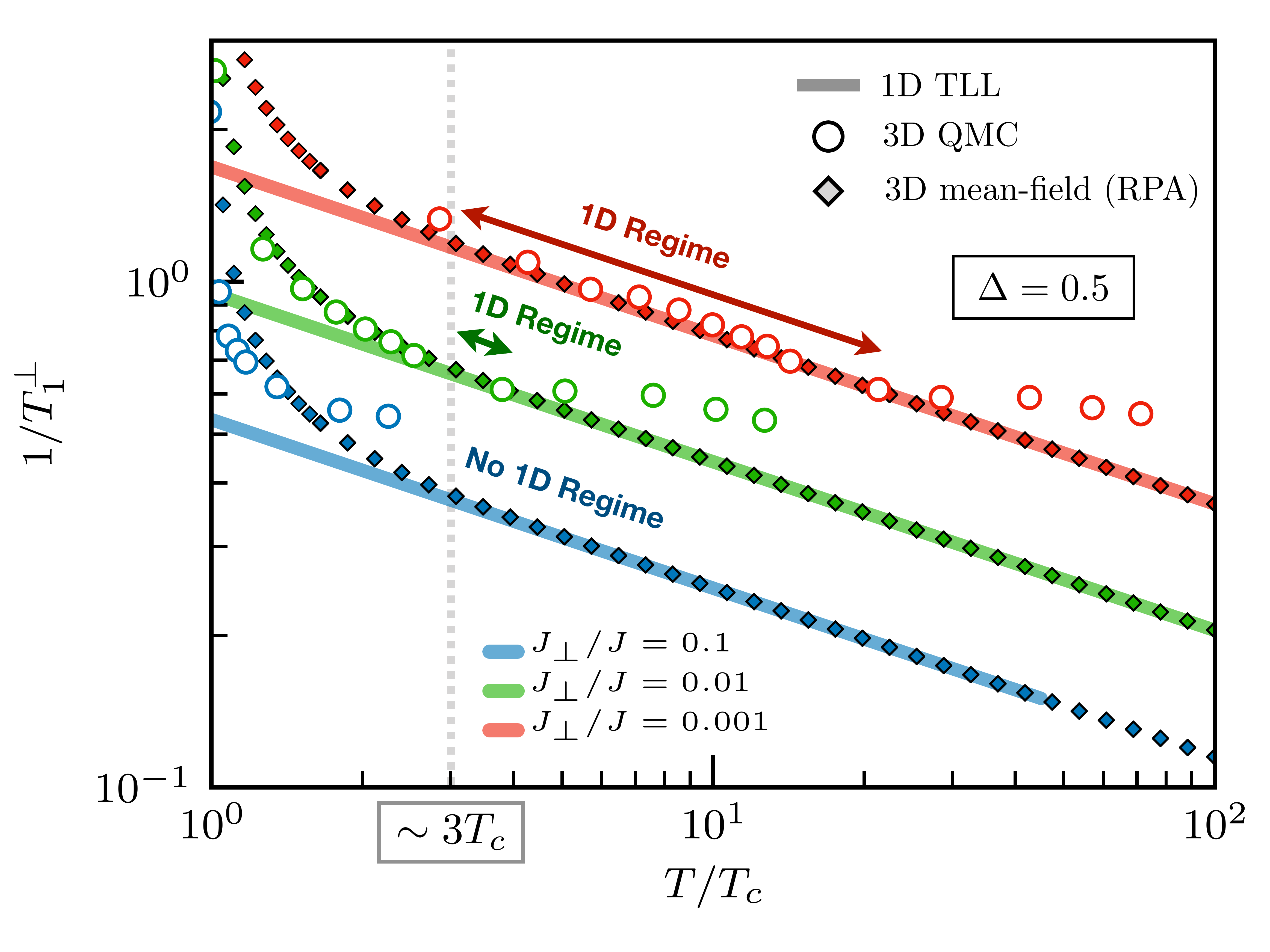}
    \caption{(Color online) Transverse component of the NMR relaxation rate $1/T_1^\perp$ defined in Eq.~\eqref{eq:t1_def2} for weakly coupled chains in $3$D with an Ising anisotropy $\Delta=0.5$ along the spatial direction $\textbf{a}$, for various transverse couplings between the chains $J_\perp/J=0.1$ (blue), $0.01$ (green) and $0.001$ (red). The temperature axis has been rescaled by the critical temperature $T_c$ of each model, respectively $T_c/J\simeq 0.224$, $0.04$ and $0.007$. The bold straight lines correspond to the purely one-dimensional TLL prediction~\eqref{eq:ll_spsm_corr}, the small diamonds to the three-dimensional mean-field (RPA) calculations~\eqref{eq:t1-qpia-tn} and the circles to quantum Monte Carlo simulations. In the latter case, the largest system available is considered for each $J_\perp/J$ value: $N=96\times 12\times 12$, $96\times 8\times 8$ and $96\times 8\times 8$, for $J_\perp/J=0.1$, $0.01$ and $0.001$ respectively. It shows that if any, the universal one-dimensional regime of the NMR relaxation rate is visible for $T\gtrsim 3T_c$.}
    \label{fig:T1_vs_TTc}
\end{figure}

\paragraph{NMR relaxation rate.---} We now turn our attention to the transverse component of the NMR relaxation rate for $T>T_c$, which is in our case directly computed from the spectral function in the limit $\omega\rightarrow 0$ according to the definition~\eqref{eq:t1_def}. Numerical simulations have been performed for weakly coupled XXZ chains $J_\perp/J=10^{-1}$, $10^{-2}$ and $10^{-3}$ with an Ising anisotropy $\Delta=0.5$ along the chain direction $\mathbf{a}$. These systems respectively develop long-range AF order below $T_c/J\simeq 0.224$, $0.04$ and $0.007$. Quantum Monte Carlo results as well as the RPA calculation of Eq.~\eqref{eq:t1-qpia-tn} and the purely one-dimensional result of Eq.~\eqref{eq:ll_spsm_corr} are plotted together in Fig.~\ref{fig:T1_vs_TTc}. In the high temperature limit, the RPA calculation gives back the purely one-dimensional prediction $\propto T^{1/2K-1}$, which becomes valid at low enough temperature $T\lesssim J/10$~\cite{dupont2016,coira2016}. For coupled chains with $J_\perp/J=0.1$, the system gets ordered above this crossover temperature preventing any one-dimensional regime. As the three-dimensional coupling is lowered, the critical temperature decreases and a $1$D regime sets up above $T_c$. Yet, the temperature should be such that $T\gg T_c$ to ensure that the transition does not spoil the universal $1$D behavior. Indeed, as we approach the transition (critical regime), the NMR relaxation rate deviates from the power-law dependence, which will be discussed thoroughly in the next section. For $J_\perp/J=10^{-2}$ and $10^{-3}$, we find that for $T\gtrsim 3T_c$ we are far enough from the transition and able to observe the $1$D regime. More precisely, we find that for $\Delta=0.5$, systems with a three-dimensional coupling $J_\perp/J<10^{-2}$ display a nonzero temperature window $T\in [3T_c, J/10]$ (assuming that $3T_c<J/10$) inside which the observation of the genuine $\propto T^{1/2K-1}$ behavior for the NMR relaxation rate is possible. We stress that in Fig.~\ref{fig:T1_vs_TTc}, there are no free parameters to adjust the different estimates.

\subsection{Critical regime}

As we approach the transition, the NMR relaxation rate is strongly enhanced, as observed in Fig.~\ref{fig:T1_vs_TTc} for $T\lesssim 3T_c$ and numerous experiments~\cite{klanjsek2015,klanjsek2015prb,jeong2013,jeong2017}. This can be understood within a scaling hypothesis since $1/T_1$ is related to a correlation function. Specifically, at the transition, we expect a divergence of both the correlation length $\xi$ and the correlation time $\tau$, linked through the relation $\tau\sim\xi^{z_t}$ with $z_t$~\footnote{This exponent is not to be mistaken with ``$z$'', the dynamic critical exponent appearing in the context of quantum phase transitions~\cite{sachdev2001}.} the dynamical exponent in the sense of real-time dynamics~\cite{susuki1977,hohenberg1977,folk2006}. Within a scaling hypothesis, the local time-dependent correlation function takes the form,
\begin{align}
    S^{\pm\mp}_{\mathbf{r}=\mathbf{0}}(t) = \xi^{2-D-\eta}\,\tilde{\mathcal{G}}\left(\xi^{1/\nu}|T-T_c|, t/\xi^{z_t}\right),
\end{align}
where $\tilde{\mathcal{G}}$ is a universal scaling function, $D$ the dimensionality of the system, $\eta$ the anomalous exponent and $\nu$ the correlation length exponent. Its Fourier transform to frequency space in the limit $\omega_0\rightarrow 0$ is the transverse component of the NMR relaxation rate~\eqref{eq:t1_def2} and simplifies to,
\begin{align}
    \frac{1}{T_1^\perp}=\xi^{2-D-\eta+{z_t}}\mathcal{G}\,\Bigl(\xi^{1/\nu}|T-T_c|\Bigr)
    \label{eq:scaling_T1}
\end{align}
where $\xi^{z_t}\mathcal{G}$ is the integral of $\tilde{\mathcal{G}}$ with $\mathcal{G}$ a universal scaling function as well.

Setting $D=3$ and using the scaling form of the correlation length $\xi\sim|T-T_c|^{-\nu}$ in Eq.~\eqref{eq:scaling_T1}, one obtains the behavior of the NMR relaxation rate when approaching the transition, $T\rightarrow T_c$,
\begin{align}
    \frac{1}{T_1^\perp}\propto |T-T_c|^{-\nu(z_t-1-\eta)},
    \label{eq:T1_at_Tc}
\end{align}
which diverges as long as $\eta<z_t-1$ since $\nu>0$.

In the classical limit, our model becomes the three-dimensional XY model, the critical dynamics of which is described by Model E of Refs.~\onlinecite{hohenberg1977,folk2006}. In model E, the non-conserved order parameter and the conserved magnetization have different dynamical exponents, respectively $z_{\phi,t}$ and $z_{m,t}$, satisfying $z_{\phi,t}+z_{m,t}=3$. In Eq.~\eqref{eq:T1_at_Tc}, we have $z_t\equiv z_{\phi,t}$ since the relaxation rate is obtained from a correlation function related to the order parameter. Two possible fixed points exist for model E dynamics~\cite{folk2006}, $z_{m,t}=z_{\phi,t}=3/2$ and $z_{m,t}<z_{\phi,t}$. Using the values of exponents obtained from numerical simulations in Ref.~\onlinecite{krech1999}, $\eta=0.035$, $\nu=0.6693$ and $z_{\phi,t}=1.62$, we find a behavior $1/T_1^\perp \sim |T-T_c|^{-0.3915}$, that should hold in the classical critical region of the transition. Alternatively, with a purely relaxational dynamics (the so-called model A)~\cite{jensen2000}, a classical dynamical exponent $z_t=2$ would be obtained, leading to $1/T_1^\perp\sim |T-T_c|^{-0.64}$. We expect that in systems where magnetization is non-conserved as a result of Dzyaloshinskii-Moriya or dipolar interactions, this model A exponent will apply. Outside this classical critical region, the mean-field exponents are recovered.

\begin{figure}[t]
    \includegraphics[width=1\columnwidth,clip]{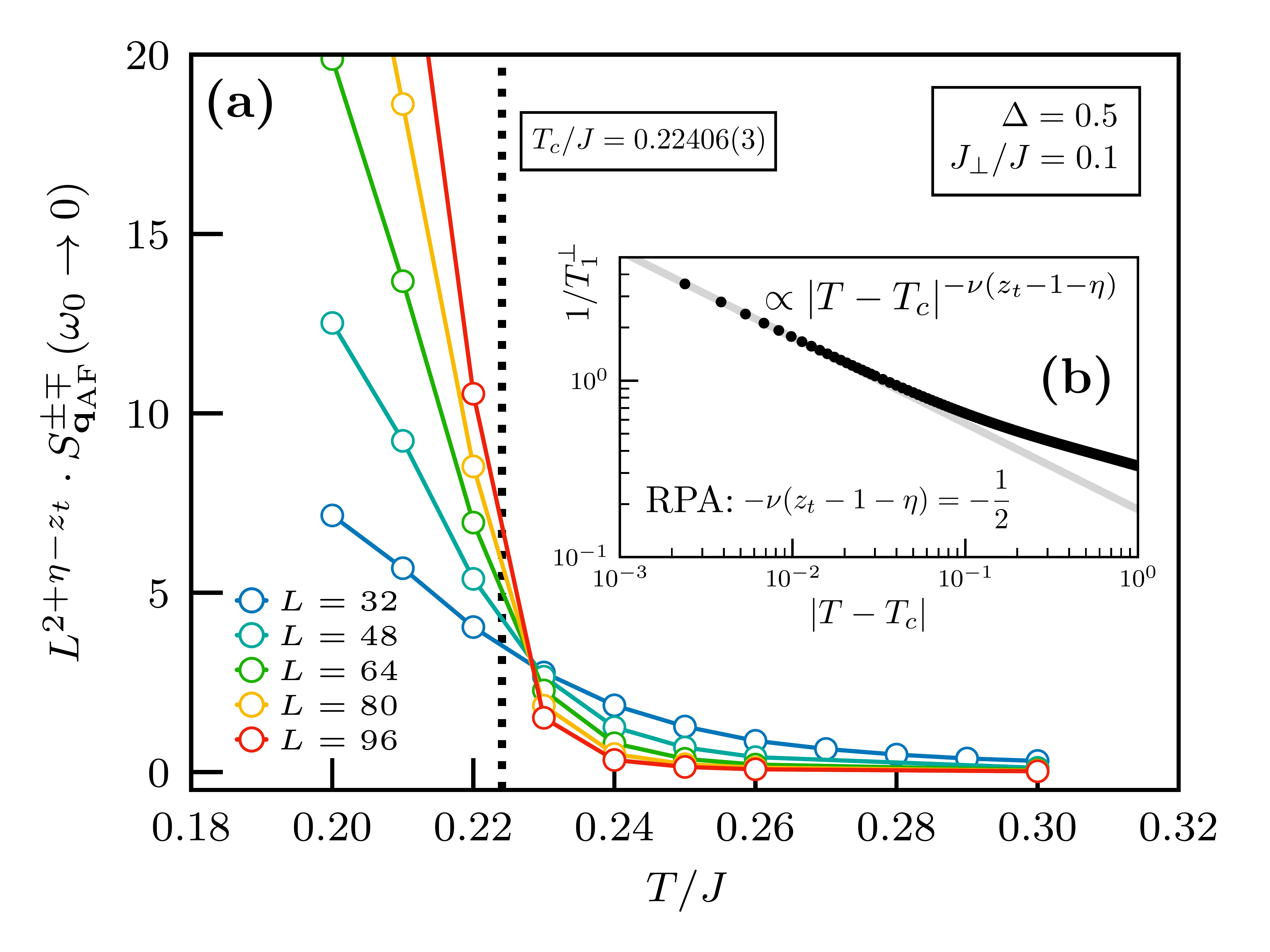}
    \caption{(Color online) {(a)} Rescaled transverse dynamical spin structure factor at the AF wave vector in the limit $\omega\rightarrow 0$ for different system sizes $N=L^3/8^2$ made of weakly coupled chains with $J_\perp/J=0.1$ and an Ising anisotropy $\Delta=0.5$ along the spatial direction $\textbf{a}$. The transverse dynamical spin structure factor has been rescaled according to the scaling law~\eqref{eq:sqAFw0_scaling}.  The anomalous critical exponent takes the value of the $3$D XY universality class $\eta=0.0381$~\cite{campostrini2006}, and $z_t=1.62$~\cite{krech1999} was considered. The crossing point for the different sizes using these exponents is close to the expected critical temperature $T_c/J\simeq 0.224$ for this system. {(b)} Transverse component of the NMR relaxation of Eq.~\eqref{eq:t1-qpia-tn} from RPA calculations versus $|T-T_c|$ with $T>T_c$. A divergence with the mean-field exponents $\nu=0.5$ and $\eta=0$ as well as $z_t=2$ is observed as $T\rightarrow T_c$ according to Eq.~\eqref{eq:T1_at_Tc}.}
    \label{fig:SqAFw0_scaling}
\end{figure}

In the vicinity of the antiferromagnetic ordering, $T\rightarrow T_c$, we can expand the denominator in the integral of Eq.~\eqref{eq:t1-qpia-tn}, and recover the mean-field behavior $1/T_1^\perp\propto |T-T_c|^{-1/2}$~\cite{giamarchi1999}, compatible with the mean-field exponents $\eta=0$, $\nu=1/2$ and $z_t=2$. This is visible in Fig.~\ref{fig:SqAFw0_scaling}\,(b) for weakly coupled chains with $J_\perp/J=0.1$ and an Ising anisotropy $\Delta=0.5$ along the spatial direction $\textbf{a}$. Regarding the subdominant contributions to the total NMR relaxation rate $1/T_1$, one can show that they do not play a role close to the transition. Indeed, $(1/T_1^\parallel)_{q_a\simeq \pi}$, is given by an integral similar to the one in Eq.~\eqref{eq:t1-qpia-tn}, but with $K\rightarrow 1/(4K)$ and $T_c\rightarrow T_c^\mathrm{Ising}$, where $T_c^\mathrm{Ising}<T_c$ is the critical temperature of a model with only Ising interchain exchange interaction. Since the enhancement of $(1/T_1^\parallel)_{q_a\simeq\pi}$ happens for $T\rightarrow  T_c^\mathrm{Ising}$, it is entirely preempted by the one of $(1/T_1^\perp)_{q_a\to\pi}$. The two terms with $q_a\simeq 0$ give contributions that are not enhanced at all in the vicinity of a transition as they remain finite for $T\to 0$, preventing a divergence when $J_\perp \ll J$. Therefore, the substitution $1/T_1^\perp\rightarrow 1/T_1$ in Eq.~\eqref{eq:T1_at_Tc} is justified since it is the dominant contribution.

A similar scaling to the transverse NMR relaxation rate of Eq.~\eqref{eq:scaling_T1} can be obtained for the transverse dynamical spin structure factor at the AF wave vector in the limit $\omega\rightarrow 0$,
\begin{equation}
    S^{\pm\mp}_{\mathbf{q}_\mathrm{AF}}(\omega_0\rightarrow 0)=\xi^{1-D-\eta+z_t}\,{{\mathcal{F}}}\Bigl(\xi^{1/\nu}|T-T_c|\Bigr),
    \label{eq:sqAFw0_scaling}
\end{equation}
with ${{\mathcal{F}}}$ a universal scaling function. At criticality, $\xi$ diverges and one can make the substitution $\xi\rightarrow L$ for a finite-size system of linear size $L$. The above scaling implies scale invariance at the critical temperature for $S^{\pm\mp}_{\mathbf{q}_\mathrm{AF}}(\omega_0\rightarrow 0)/L^{1-D-\eta+z_t}$.

We plot in Fig.~\ref{fig:SqAFw0_scaling}\,(a) setting $D=3$ and $z_t=1.62$ (see previous discussion) and using the $3$D XY universality class value of the exponent $\eta=0.0381$~\cite{campostrini2006}: it is noteworthy that the different curves show a crossing point close to the critical temperature $T_c/J\simeq 0.224$ of the system made of weakly coupled chains with $J_\perp/J=0.1$ and an Ising anisotropy $\Delta=0.5$ along the spatial direction $\textbf{a}$. However, the crossing is not extremely accurate, which could be related either to the numerical value of $z_t$ or more probably to the analytic continuation procedure. Similarly, we are unable to get the universal scaling function as a rescaling of the $x$ axis by $T\rightarrow (T-T_c)L^{1/\nu}$ with $\nu=0.6717$~\cite{burovski2006,campostrini2006,beach2005} (the $3$D XY universality class value of the correlation length exponent) does not provide a satisfactory collapse of our data. The inability to properly estimate error bars of analytically continued data is partially to blame, but more importantly the diverging value of $S_{\mathbf{q}_\mathrm{AF}}(\omega_0\rightarrow 0)$ below the critical temperature is not accurately evaluated. It is known that analytic continuation has troubles to capture sharp peaks such as $\delta$ or quasi-$\delta$ contributions in spectral functions like the one present in $S_{\mathbf{q}_\mathrm{AF}}(\omega)$ as $\omega\rightarrow 0$ below the critical temperature.

\subsection{Ordered phase}

As discussed in the previous section, the strong enhancement of the $1/T_1$ when approaching $T_c$ is understood within a scaling hypothesis, provided $z_t+2-D-\eta>0$. In the ordered phase, a linear dependence of the $1/T_1$ with the temperature is predicted~\cite{giamarchi1999} due to spin-waves contribution but has never been observed experimentally so far. Instead, a stronger suppression of the NMR relaxation rate is reported $\propto T^\alpha$ with $\alpha\simeq 4-5$, as in the two-leg spin-$1/2$ ladder Cu$_2$(C$_5$H$_{12}$N$_2$)$_2$Cl$_4$ compound~\cite{mayaffre2000}, DTN~\cite{blinderthesis} and DIMPY~\cite{jeong2017}. We show in this section that the linear spin-waves contribution should manifest only at low temperature and discuss the meaning of the strongly suppressed $1/T_1$ close to $T_c$ by looking at its different momenta components.

\subsubsection{Close to the transition}

\begin{figure}[t]
    \includegraphics[width=1\columnwidth,clip]{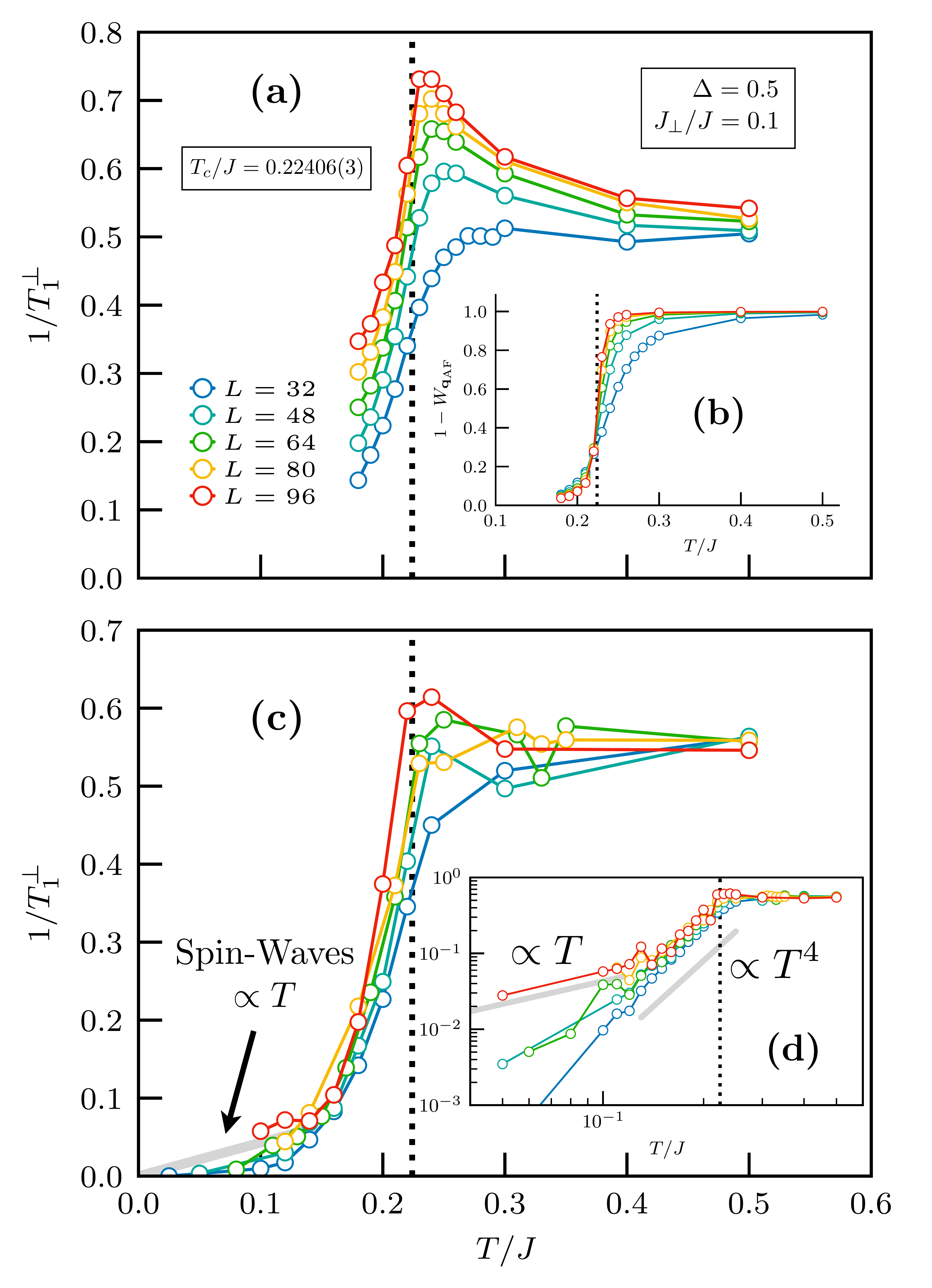}
    \caption{(Color online) Transverse component of the NMR relaxation rate $1/T_1^\perp$ Eq.~\eqref{eq:T1noqAF_def} where the $\mathbf{q}_\mathrm{AF}$ component has been removed (see discussions in main text) for weakly coupled chains in $3$D with an Ising anisotropy $\Delta=0.5$ along the spatial direction $\textbf{a}$. It has been computed numerically using QMC+SAC on different system sizes $N=L^3/8^2$ with an interchain coupling $J_\perp/J=0.1$, leading to a critical temperature $T_c/J\simeq 0.224$ (vertical dotted line). In panel {(a)} and its inset {(b)}, the stochastic analytic continuation has been performed independently on all $\mathbf{q}$ components of $S_\mathbf{q}(\omega_0)$ and summed thereafter to obtain the NMR relaxation rate. In panels {(c)} and {(d)}, the sum over $\mathbf{q}$ of the imaginary time QMC data is performed before doing the analytic continuation. The inset {(b)} shows the relative weight of the $\mathbf{q}\neq\mathbf{q}_\mathrm{AF}$ components in the $1/T_1$ as a function of the temperature with $W_{\mathbf{q}_\mathrm{AF}}$ defined in Eq.~\eqref{eq:T1_qAFweight}. The inset {(d)} is the same as {(c)} in log-log scale where the power-law compatible with $\propto T^4$ can be observed. At lower temperature, the spin-waves contribution of the $1/T_1\propto T$ is plotted with the prefactor computed by SCHA in Eq.~\eqref{eq:sw-nmr-t1} (with no free parameter).}
    \label{fig:T1_vs_T}
\end{figure}

From the definition~\eqref{eq:t1_def}, the NMR relaxation rate can be expressed as a sum over all momenta $\mathbf{q}$ of the dynamical spin structure factor $S_\mathbf{q}(\omega_0)$ at the NMR frequency. We show in Fig.~\ref{fig:T1_vs_T}\,(b) the relative weight versus temperature of the AF momentum $\mathbf{q}_\mathrm{AF}$ compared to all the others by defining,
\begin{align}
    \label{eq:T1_qAFweight}
    W_{\mathbf{q}_\mathrm{AF}} = \frac{S_{\mathbf{q}_\mathrm{AF}}(\omega_0)}{\sum_\mathbf{q}S_\mathbf{q}(\omega_0)}.
\end{align}
The AF wavevector clearly dominates below the critical temperature as hinted by Fig.~\ref{fig:SqAFw0_scaling}\,(a) showing its diverging behavior below $T_c$ (the exponent $2+\eta-z_t$ is close to zero and the multiplicative factor $L^{2+\eta-z_t}$ on the $y$ axis of order one). As mentioned a couple of times, there is experimentally no divergence of the $1/T_1$ below $T_c$ but a strong suppression. This can only mean that the sharp AF contribution at low frequency $\omega\rightarrow 0$ is not captured in the ordered phase, which can be explained by the finiteness of the NMR frequency $\omega_0$. To avoid a specific dependence on the NMR frequency of the relaxation rate, we make a new definition removing the $\mathbf{q}_\mathrm{AF}$ contribution,
\begin{equation}
    \frac{1}{T_1^\perp}=\sum_{\mathbf{q}\neq\mathbf{q}_\mathrm{AF}}S^{\pm\mp}_\mathbf{q}(\omega_0\rightarrow 0).
    \label{eq:T1noqAF_def}
\end{equation}
The regular contribution, if any, at low frequency of the AF component is also dismissed in this definition but should not contribute more than any other wavevector and only induce an error of order $1/N$, with $N$ the number of spins (or equivalently the number of terms in the sum).

Focusing on weakly coupled chains $J_\perp/J=0.1$ with  Ising anisotropy $\Delta=0.5$ along the spatial direction $\textbf{a}$, we plot in Fig.~\ref{fig:T1_vs_T}\,(a) the NMR relaxation rate from the definition~\eqref{eq:T1noqAF_def} with no AF contribution. The stochastic analytic continuation has been performed independently on the $N-1$ dynamical spin structure factors in momentum space and summed thereafter~\footnote{Here we exploit the tetragonal structure of the studied Hamiltonian~\eqref{eq:hamiltonian} to only perform analytic continuation on $\mathbf{q}$ points of the irreducible BZ, see Fig.~\ref{fig:lattice_bz}. In the thermodynamic limit, this reduces the computational cost by a non-negligible factor $\sim 16$ (from $5$ to $10$ for the finite lattices studied in the present work).}. The $1/T_1$ is very little affected in the disordered phase from the $\mathbf{q}_\mathrm{AF}$ component removal. It still displays a diverging behavior when approaching the critical temperature (the maximum value increases with system size) and the position of the maximum gets closer and closer to the actual value of $T_c$ as the system size is increased. In the ordered phase, the NMR relaxation rate is suppressed for each one of the sizes but still growing with system size $N$. This is undoubtedly a technical artifact of the stochastic analytic continuation: it is not able to resolve accurately the very small contributions of the different $\mathbf{q}$ points which are all added up at the end. This can be seen as the sum of positive-definite (since a spectral function is) ``numerical noise''; note the hundreds to thousands contributions added up.

In an attempt to overcome this issue, we first perform the sum of the imaginary time data resulting from the quantum Monte Carlo simulations, except for $\mathbf{q}_\mathrm{AF}$ component, and then run a single analytic continuation. The result is shown in Fig.~\ref{fig:T1_vs_T}\,(c) for the same system as panel (a), and is visually not as smooth as the first panel. The high-temperature regime is not as well-captured as before with no precise maximum defined at the transition. On the contrary, in the ordered regime, the NMR relaxation rate seems more or less independent of the system size, a good indicator since it is a local probe. The same data are shown in Fig.~\ref{fig:T1_vs_T}\,(d) in log-log scale. It becomes increasingly difficult at low temperature to collect an accurate estimate of the NMR relaxation rate which becomes exceedingly small. Nonetheless, we are able to observe a strong suppression below $T_c$, compatible with a power-law dependence $1/T_1^\perp\propto T^4$ as experimentally measured.

\subsubsection{Spin-waves contribution at low temperature}

\begin{figure*}[t]
    \includegraphics[width=2\columnwidth,clip]{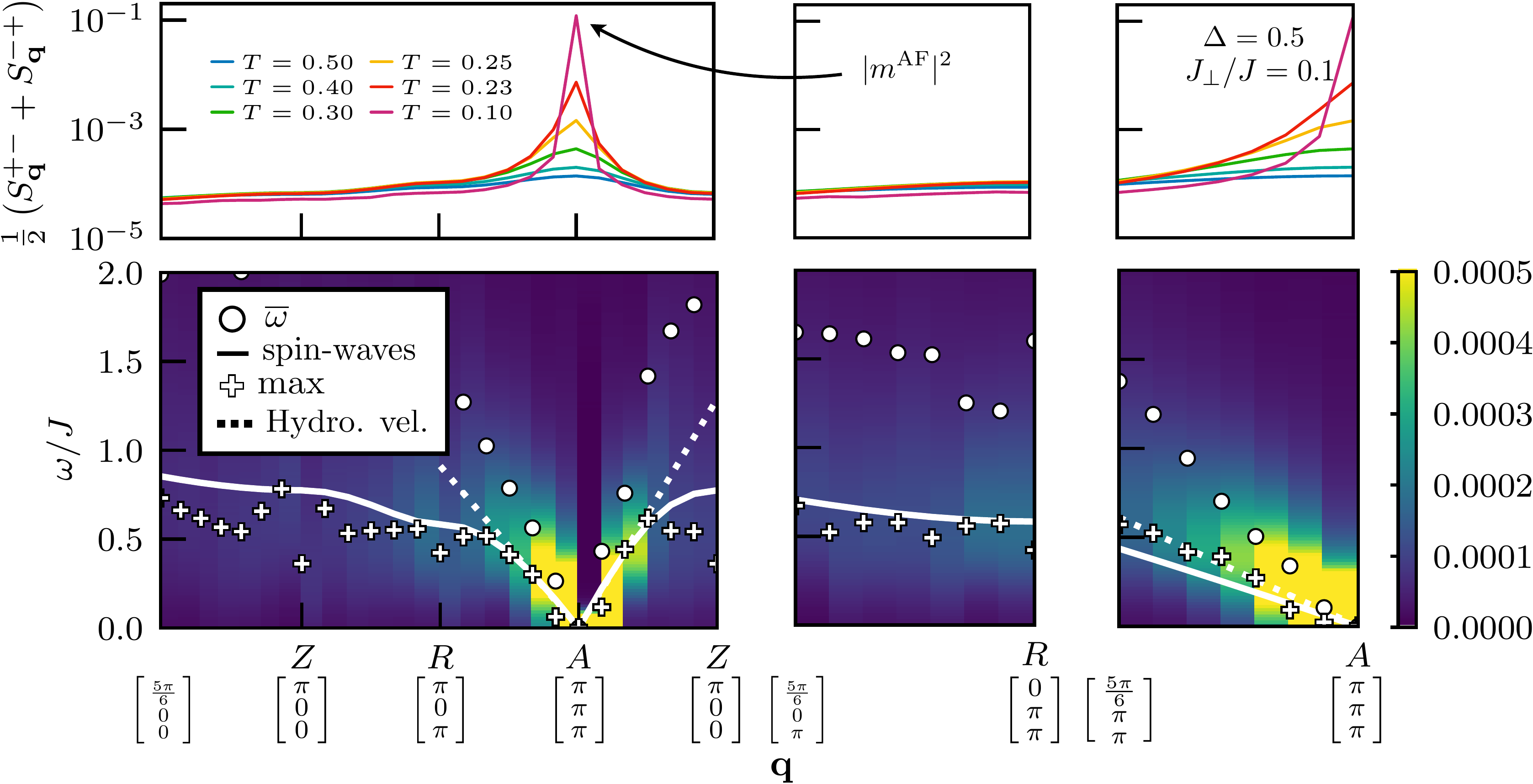}
    \caption{(Color online) Lower panels: transverse inelastic neutron scattering intensity $I^\perp$ for weakly coupled spin chains in three dimensions with $J_\perp/J=0.1$ and an Ising anisotropy $\Delta=0.5$ along the spatial $\textbf{a}$ direction. The $\mathbf{q}$ points follow the high symmetry lines of the BZ of Fig.~\ref{fig:lattice_bz}\,(b), focusing on regions where the spectral weight is the more significant. The temperature of the system is $T=0.1J$, below the critical temperature $T_c/J\simeq 0.224$, explaining why all the spectral weight is located at the AF wave vector. The white dot symbols show the first moment of the spectrum. The straight white line is the spin-waves dispersion relation $\omega_\mathrm{sw}(\mathbf{q})$ with a gapless mode at the AF wave vector. The plus symbols correspond to the maximum of intensity in the spectrum at a given $\mathbf{q}$ point. The dotted white lines around the AF wavevector show the linear dispersion relation around $\mathbf{q}_\mathrm{AF}$ of the SW spectrum but with corrected (hydrodynamic) velocities, compared to the bare spin-waves ones, see text. For visibility, the color intensity has been saturated to $0.0005$.  The upper panels correspond to the static structure factor whose value at $A\equiv\mathbf{q}_\mathrm{AF}$ is the modulus square of the order parameter which clearly develops for $T<T_c$ (a careful finite-size scaling analysis would need to be performed in order to obtain the order parameter value in the thermodynamic limit $N\rightarrow\infty$). The data are from quantum Monte Carlo simulations on the largest available system of size $N=96\times 12 \times 12=13\,824$ spins.}
    \label{fig:colormap_3d_below_tc}
\end{figure*}

Deep in the ordered phase at zero temperature, the spin-waves (SW) dispersion relation can be obtained by treating semi-classically the Hamiltonian~\eqref{eq:hamiltonian}. The idea is to first make a rotation of the spin operators in order to align the quantization axis with the classical order along the $x$ direction~\cite{coletta2012}. Then, the Dyson-Maleev representation of the $S=1/2$ operators is introduced and only quadratic terms are kept. In this representation, the truncated Hamiltonian is diagonalized through a Bogoliubov transformation with the SW excitation spectrum given by $\omega_\mathrm{sw}(\mathbf{q})=\sqrt{A_\mathbf{q}^2-B_\mathbf{q}^2}$, where
\begin{align}
    A_\mathbf{q} = 2J_\perp + J + \left(\frac{\Delta-1}{2}\right)\cos\left(\pi-q_a\right),
\end{align}
and
\begin{multline}
    B_\mathbf{q} = J\left(\frac{\Delta+1}{2}\right)\cos\left(\pi-q_a\right)\nonumber\\ +J_\perp\Bigl[\cos\left(\pi-q_b\right)+\cos\left(\pi-q_c\right)\Bigr],
\end{multline}
with a zero mode at the AF wavevector $(\pi,\pi,\pi)$. Expanding the cosines close to the antiferromagnetic wave vector, we obtain a linear dispersion relation $\omega_\mathrm{sw}(\mathbf{q}\rightarrow\mathbf{q}_\mathrm{AF})\sim v_\mathrm{sw}^\nu|\mathbf{q}-\mathbf{q}_\mathrm{AF}|$ with $v_\mathrm{sw}^\nu$ the SW velocity, which depends on the direction $\nu\in[a,b,c]$ of the Brillouin zone for an anisotropic system,
\begin{equation}
    v_\mathrm{sw}^\nu=\sqrt{J_\nu\left[J\left(\frac{\Delta+1}{2}\right)+2J_\perp\right]},
    \label{eq:sw_velocity}
\end{equation}
with $J_a\equiv J$ and $J_{b,c}\equiv J_\perp$. We plot in Fig.~\ref{fig:colormap_3d_below_tc} the spectral function in the ordered phase of weakly coupled chains with $J_\perp/J=0.1$ and an Ising anisotropy $\Delta=0.5$ along the spatial $\textbf{a}$ direction. As expected in the ordered phase, the maximum of intensity is located at $\textbf{q}_\mathrm{AF}$ with a zero mode. We also compute the first moment of the spectral function (white dots) and display the position of the maximum of intensity (plus symbols). We only focus on regions of the BZ close to the AF wavevector, where the spectral weight is the more significant to be reliable (note that we have the color intensity has been saturated for visibility). The spin-waves dispersion relation $\omega_\mathrm{sw}(\mathbf{q})$ derived above is also shown (straight line) and overlaps pretty well with the maximum of intensity, which seems more relevant than the first moment here. The linear dispersion above the ground state around the antiferromagnetic wave vector is overall well-captured, with the linear slope given by the SW velocity of Eq.~\eqref{eq:sw_velocity}. This is especially true in the transverse $RA$ and $AZ$ directions but the maximum of intensity deviates from the SW dispersion relation along the chain direction as observed in the right panel. This is certainly due to the linear spin-wave approximation restricted to $\mathcal{O}(1/S)$ corrections. To go further, a well-known way to extract the velocity in an antiferromagnet is to use the analog of a hydrodynamic relationship relating the velocity to the spin stiffness and the susceptibility~\cite{halperin1969,sen2015},
\begin{equation}
    v^\nu_\mathrm{hydro} = \sqrt{\frac{\rho_s^\nu}{\chi}},
    \label{eq:velocity_hydro}
\end{equation}
where $\chi$ is the magnetic susceptibility and $\rho_s^\nu$ the spin stiffness in the $\nu\in[a,b,c]$ spatial direction. We computed both quantities for the system studied here, performing a careful finite-size scaling analysis ($N\rightarrow\infty$) and making sure that we were probing the ground state by being at sufficiently low temperature. Our final estimates are $v^a_\mathrm{hydro}=1.334(6)$ and $v^{b,c}_\mathrm{hydro}=0.29(1)$, plotted as dotted lines in Fig.~\eqref{fig:colormap_3d_below_tc}. The correction is almost invisible in the transverse directions but provides a better overlap to the maximum of intensity along the chain direction.

The Green's function of the bosonized field $\theta_{\mathbf{r}_\perp}(x)$ obtained from SCHA is, in Fourier representation,
\begin{multline}
    \label{eq:green-varia}
    G(\mathbf{q},\omega) = \Biggl\{\frac{K}{\pi u} \Bigl[\omega_n^2 + (u
    q_a)^2\Bigr]\\
    +\kappa(T)\Bigl[2-\cos(\mathbf{q}\cdot \mathbf{b}) - \cos(\mathbf{q}\cdot \mathbf{c})\Bigr]\Biggr\}^{-1},
\end{multline}
where $\kappa(T)$ is determined by a self-consistent equation (see App.~\ref{app:scha-solution}). The resulting spin-wave velocity in the transverse direction is:
\begin{equation}
    \label{eq:sw-velocity}
    \Bigl(v_\mathrm{sw}^{b,c}\Bigr)^2 = \frac{\pi \kappa(T) u}{2 K},
\end{equation}
and we obtain (see App.~\ref{app:scha-response}) the low temperature spin-wave contribution as:
\begin{equation}
    \label{eq:sw-nmr-t1}
    \frac{1}{T_1^\perp} = \Biggl[\frac{m^\mathrm{AF}(T)}{v_\mathrm{sw}^{b,c}}\Biggr]^2 \frac{T}{4K},
\end{equation}
where $m^\mathrm{AF}(T)$ is the expectation value of the order parameter, which saturates quickly to the zero temperature value. At very low temperatures, we have $1/T_1^\perp \sim T/(JJ_\perp)$, while in general for $T<T_c$, $1/T_1^\perp \sim T\varphi(T/T_c)$, $\varphi(T/T_c)$ representing a scaling function deduced from SCHA.  The SCHA gives a linear behavior of $1/T_1^\perp$ till $T\simeq 0.6 T_c$, with a modest superlinear increase above this temperature. Since the SCHA is known to underestimate contributions from topological excitations~\cite{cazalilla2006,you2012}, that too weak
enhancement near the critical region is not really surprising. The results that are likely to be reliable are the ones at low temperature, where the relaxation rate is linear.

\section{Summary and conclusions}\label{sec:summary_conclusions}

We have provided an analytical and numerical study of the full temperature behavior of spin dynamical spectral functions for quasi-one-dimensional quantum antiferromagnets. Since our motivation comes from coupled-ladder materials or coupled chains in finite magnetic field, we have chosen to focus on U(1) symmetric models which exhibit a spontaneous staggered magnetization in the XY plane below a three-dimensional critical temperature $T_c$. Our main findings regarding the NMR $1/T_1$ relaxation rate are plotted in Fig.~\ref{fig:T1_regimes} and we can distinguish several temperature ranges:
\begin{itemize}
    \item $3T_c \lesssim T\lesssim 0.1J$: this is a universal regime where the interchain couplings are irrelevant so that low-energy properties are well described using a Tomonaga-Luttinger framework.
    \item Critical regime close to $T_c$: because of the finite-temperature transition (continuous, in the 3D XY universality class), the $1/T_1$ rate diverges due to the strong increase of the spin structure factor at the AF wave vector. This has been obtained both in our mean-field approach (using RPA) and in our numerical data. Some uncertainty remains about the precise value of the critical exponent, which involves the real-time dynamic exponent $z_t$ that has not been measured precisely in such systems. It would be an interesting prospect to determine its value using the real-time dynamics of the order parameter.
    \item In the ordered phase, most of the spectral weight is contained in a zero-frequency delta peak at the AF wave vector, but because of the finiteness of the NMR frequency, it does not contribute to $1/T_1$. As a result, there is a large reduction of this quantity when the temperature decreases below $T_c$. Our numerical analysis gives indication of a power-law behavior $1/T_1 \sim T^4$ (note that this exponent is fitted on a small temperature window), which is compatible indeed with several experimental observations. At much lower temperature, the ordered phase can be well described using spin wave analysis, in agreement with our numerical data on the spin dynamical correlations, and the $1/T_1$ is predicted to have a linear dependence in $T$. Quite interestingly, we have also used the so-called self-consistent harmonic approximation (SCHA) to compute the prefactor of this linear behavior, and this approximation is expected to be very good at low enough temperature. This is particularly important since the low-temperature regime is the most difficult to tackle numerically.
\end{itemize}

Overall, by combining analytical and numerical approaches, we have described a very rich and non-trivial temperature-dependence of the NMR relaxation rate, with similar findings in the full spectral functions, which can provide useful information when analyzing experimental data (NMR or INS). For instance, we have pointed out that for some materials, the 1D universal regime might not be accessible so that an accurate determination of the Luttinger parameter $K$ (as often done experimentally from fitting $1/T_1$ power-law) is not possible.

In the near future, we plan to investigate the fully $\mathrm{SU}(2)$ symmetric case, which transition is in a different universality class. This case is more difficult to tackle analytically since the real-space response functions of the single chain present logarithmic corrections and no explicit expression of their Fourier transform is known. This prevents the application of RPA methods to obtain the behavior of the response functions above the transition, unless logarithmic corrections are neglected. In the low temperature phase, the situation is worse since applying the self-consistent harmonic approximation would violate the $\mathrm{SU}(2)$ symmetry, yielding incorrect results, in particular for the Goldstone modes. More sophisticated analytical approaches, that can fully preserve the symmetry will have to be developed.

As a last remark, it would be also interesting to consider the case of coupled chains in two-dimensions~\cite{raczkowski2013}. However, the situation in this case is even less favorable for an analytical approach. Indeed, the Mermin-Wagner theorem prohibits the existence of long-range ordering at any positive temperature~\cite{mermin1966}. For chains forming a rectangular lattice, the low temperature phase has only quasi-long range order~\cite{berezinskii1971,kosterlitz1973,kosterlitz1974} until the Berezinskii-Kosterlitz-Thouless (BKT) transition~\cite{berezinskii1971,kosterlitz1973,kosterlitz1974} where short range order sets in. In such situation, mean field theory breaks down since the gaussian fluctuations around the saddle-point cannot be controlled~\cite{lebellac1992}.  However, SCHA~\cite{you2012} correctly reproduces the quasi-long range ordered phase, and can be used to predict the BKT transition~\cite{berezinskii1971,kosterlitz1973,kosterlitz1974} temperature. But it incorrectly predicts a first order transition~\cite{donohue2001}, indicating its breakdown at temperatures of the order of the BKT transition temperature. The SCHA might thus be applicable, as in the $3$D case, to the calculation of the NMR relaxation rate near zero temperature.

\acknowledgments{The authors are grateful to M. Horvati\'c for valuable discussions. E. O. would like to thank T. Giamarchi and R. Citro for useful discussions regarding the computation of the NMR relaxation rate $1/T_1$ using RPA. M. D. and S. C. acknowledge A. W. Sandvik for many enlightening and helpful discussions about analytic continuation. We acknowledge support of the French ANR program BOLODISS (Grant No. ANR-14-CE32-0018) and R\'egion Midi-Pyr\'en\'ees. The numerical work was performed using HPC resources from GENCI (Grant No. x2017050225, A0010500225 and A0030500225) and CALMIP. The calculations involving quantum Monte Carlo were partly based on the ALPS library~\cite{bauer2011}.

\clearpage
\appendix

\section{Expression of correlation amplitudes }\label{app:amplitudes}

The correlation amplitudes in Eqs.~\eqref{eq:chain-t0-pm} and~\eqref{eq:chain-t0-zz} are expressed~\cite{lukyanov1999,lukyanov2003,hikihara2004} as a function of the TLL parameter $K>1/2$ as:
\begin{widetext}
    \begin{eqnarray}
    \label{eq:amplitude-aperp}
    && A_\perp=\left(\frac{K}{2K-1}\right)^2 \left[\frac{\Gamma\left(\frac{1}{4K-2}\right)}{2\sqrt{\pi}\Gamma\left(\frac{K}{2K-1}\right)}\right]^{\frac 1 {2K}} \exp\left\{-\int_0^{+\infty}\frac{\mathrm{d}t}t \left[\frac{\sinh\left(\frac t{2K}\right)}{\sinh t \cosh \left(\frac{2K-1}{2K} t\right)} -\frac{e^{-2t}}{2K}\right]\right\}\\
    \label{eq:amplitude-aperp-tilde}
    &&\tilde{A}_\perp=\frac{4K^2}{2K-1} \left[\frac{\Gamma\left(\frac{1}{4K-2}\right)}{2\sqrt{\pi}\Gamma\left(\frac{K}{2K-1}\right)}\right]^{2K+ \frac 1 {2K}}\exp\Biggl\{-\int_0^{+\infty}\frac{\mathrm{d}t} t \Biggl[\frac{\cosh \left(\frac t K\right) e^{-2t}-1}{2\sinh \left(\frac t {2K}\right) \sinh t \cosh \left(\frac{2K-1}{2K} t\right)}\nonumber\\
    &&\qquad\qquad\qquad\qquad\qquad\qquad\qquad\qquad\qquad\qquad\qquad\qquad\qquad+\frac 1 {\sinh \left(\frac t {2K}\right)} -\left(2K+\frac 1{2K}\right)e^{-2t} \Biggr] \Biggr\}\\
    \label{eq:amplitude-aparallel}
    && A_\parallel=\frac 2 {\pi^2} \left[\frac{\Gamma\left(\frac{1}{4K-2}\right)}{2\sqrt{\pi}\Gamma\left(\frac{K}{2K-1}\right)}\right]^{2K}\exp\left\{\int_0^{+\infty} \frac{\mathrm{d}t}t \left[\frac{\sinh\left(\frac{1-K}{K} t\right)}{\sinh\left(\frac t {2K}\right) \cosh\left(\frac{2K-1}{2K} t\right)} -(2-2K)e^{-2t}\right] \right\}
    \end{eqnarray}
\end{widetext}

\section{Derivation of the square lattice local Green's function}\label{app:integral}
To derive the integral
\begin{equation}
  \label{eq:watson-like}
  I(z)=\int \frac{\mathrm{d}q_x \mathrm{d}q_y}{(2\pi)^2} \frac 1 {(z- \cos q_x - \cos
    q_y)^2},
\end{equation}
used to express~\eqref{eq:t1-qpia-tn}, we first introduce~\cite{morita1971}:
\begin{equation}
    \label{eq:green-2d}
    G(z)=\int_{-\pi}^\pi \frac{\mathrm{d}q_x}{2\pi} \int_{-\pi}^\pi\frac{\mathrm{d}q_y}{2\pi} \frac{1}{z-\cos q_x -\cos q_y},
\end{equation}
which can be interpreted as the local Green's function of a free
electron on a two-dimensional square lattice. We have
$I(z)=-\mathrm{d}G(z)/\mathrm{d}z$ so we only need~\eqref{eq:green-2d}.
We can easily show that for $z>2$,
\begin{equation}
  G(z)=\int_{-\pi}^\pi \frac{\mathrm{d}q_x}{2\pi} \frac{1}{\sqrt{(z-\cos q_x)^2
      -1}}.
\end{equation}
With the change of variables $v=(1+\cos q_x)/2$ we can rewrite:
\begin{equation}
  G(z)=\int_0^1 \frac{\mathrm{d}v}{\sqrt{v(1-v)(z+2-2v)(z-2v)}},
\end{equation}
which is expressible in terms of a complete  elliptic integral of the first kind~\cite{abramowitz1972} as:
\begin{equation}
    G(z)=\frac{2}{\pi z}\mathrm{K}\left(\frac 4 {z^2}\right).
\end{equation}
By differentiation, we finally find
\begin{eqnarray}
    \label{eq:watson-like-solution}
    I(z)=\frac{2}{\pi (z^2-4)} \mathrm{E}\left(\frac 4 {z^2}\right),
\end{eqnarray}
where $\mathrm{E}(x)$ is a complete  elliptic integral of the secondkind~\cite{abramowitz1972}.

\section{solution of SCHA equations}\label{app:scha-solution}
The self-consistency condition is:
\begin{equation}
  \label{eq:scha-eqn}
  \kappa(T) = J_\perp A_\perp e^{-\frac 1 2 \langle
    (\theta({\mathbf{r_\perp+b}})-\theta(\mathbf{r_\perp}))^2\rangle},
\end{equation}
where averages are taken with the Green's
function~\eqref{eq:green-varia}. Introducing a dimensionless parameter
\begin{equation}
    \gamma(T)=\frac{\kappa(T)}{4\pi u K},
\end{equation}
we find, for $T=0$, that the self-consistency Eq.~\eqref{eq:scha-eqn} is satisfied for
\begin{equation}\label{eq:scha-3d-coupling}
  \gamma(T=0) = \left(\frac{J_\perp A_\perp}{\pi u K}
    e^{\frac{\mathcal{C}_3}{8K}} \right)^{\frac{4K}{4K-1}},
\end{equation}
with
\begin{equation}
  \mathcal{C}_3=\int \frac{\mathrm{d}\theta_1 \mathrm{d}\theta_2}{(2\pi)^2} [2-\cos
  \theta_1-\cos
  \theta_2] \ln [2-\cos
  \theta_1-\cos
  \theta_2],
\end{equation}
which allows us to obtain:
\begin{equation}
  \label{eq:scha3d-gs-sx}
  \langle e^{i \theta} \rangle (T=0) = \left(\frac{J_\perp A_\perp
      }{\pi u K} e^{\frac{\mathcal{C}_3}{8K}}\right)^{\frac 1 {8K-2}}
  e^{\frac {\mathcal{C}_4}
    {16K}} ,
\end{equation}
with the same power-law scaling as in chain mean-field
theory~\cite{klanjsek2008} and:
\begin{equation}
  \mathcal{C}_4= \int \frac{\mathrm{d}\theta_1\mathrm{d}\theta_2}{(2\pi)^2} \ln (2-\cos \theta_1 -\cos
    \theta_2).
\end{equation}
For $T>0$, the self-consistency Eq.~\eqref{eq:scha-eqn} becomes:
\begin{eqnarray}
    &&\frac{\zeta(4)}{8K-2} \left(\frac{\pi T}{\sqrt{2} u}\right)^4 \frac 1 {\gamma^2(0)} =
    \frac{\zeta(4)}{8K-2} \left(\frac{\pi T}{\sqrt{2} u}\right)^4\nonumber\\
    &&\qquad\times\frac{1}{\gamma^2(T)}\exp\left[-\frac{\zeta(4)}{8K-2}  \left(\frac{\pi  T}{\sqrt{2} u}\right)^4 \frac 1 {\gamma^2(T)}\right],
\end{eqnarray}
which, provided
\begin{equation}\label{eq:tcscha}
 T\le T_c^\mathrm{SCHA} = \frac{u\sqrt{2}}{\pi } \left(\frac{8K-2}{e
   \zeta(4)}\right)^{1/4} \left(\frac{J_\perp A_\perp }{\pi u K}
   e^{\frac{\mathcal{C}_3}{8K}} \right)^{\frac{2K}{4K-1}},
\end{equation}
has a solution that can be expressed using the Lambert $W_p$ function~\cite{olver2010} as
\begin{equation}
  \gamma^2(T)=-\gamma^2(0) \frac{
   \frac 1 e \left(\frac{T}{T_c^\mathrm{SCHA}}\right)^4 }{W_p\left[-\frac{1} e \left(\frac{T}{T_c^\mathrm{SCHA}}\right)^4 \right]}.
\end{equation}

Using the asymptotic expansion~\cite{olver2010} of $W_p(z \to 0)$, one can check the continuity for $T\to 0$. When approaching the critical temperature defined by Eq.~\eqref{eq:tcscha} from below, $\gamma^2(T\to T_c^-) = \gamma^2(0)/e$ while $\gamma(T\to T_c^+)=0$, i.e., $\gamma$ is discontinuous at the transition. The order parameter behaves as:
\begin{eqnarray}\label{eq:op-scha-tpos}
  \langle e^{i\theta} \rangle(T) &=& \left[\gamma(T)\right]^\frac{1}{8K}\exp
  \left[-\frac{\pi}{48K\gamma(T)} \left(\frac{T}{\pi u}\right)^2\right]e^{\frac{\mathcal{C}_4} {16K}} \nonumber \\
  &=&  \langle e^{i\theta} \rangle(T=0) \left(\frac{-\frac 1 e \left(\frac{T}{T_c^\mathrm{SCHA}}\right)^4 }{W_p\left[-\frac{1}e \left(\frac{T}{T_c^\mathrm{SCHA}}\right)^4 \right]}\right)^{\frac 1
      {8K}}\\ && \times \exp\left\{-\frac{(4K-1)}{12\pi^3 K} \sqrt{-W_p\left[-\frac{1}
      e \left(\frac{T}{T_c^\mathrm{SCHA}}\right)^4 \right]}\right\}\nonumber
\end{eqnarray}
and is also discontinuous at the transition. This shows that the SCHA method is applicable only well below the critical temperature~\cite{donohue2001,cazalilla2006}. According to Eq.~\eqref{eq:op-scha-tpos}, the order parameter obeys a scaling law as a function of $T/T_c^\mathrm{SCHA}$.

\section{Local response functions deduced from the SCHA}\label{app:scha-response}

The local response function:
\begin{equation}
    \label{eq:chiloc-def}
    \chi_{\mathrm{loc.}}^{\pm}(t)=i \Theta(t) \Bigl\langle [S^+_{\mathrm{r}}(t),S^+_{\mathrm{r}}(0)]\Bigr\rangle,
\end{equation}
is used to calculate the transverse component of the NMR relaxation raten $1/T_1^\perp$. Within the SCHA formalism, it can be expressed as:
\begin{eqnarray}\label{eq:chiloc-scha}
    \chi_{\mathrm{loc.}}^{\pm}(t,T)&=&i \Theta(t) A_\perp e^{-\frac 1 2 \langle (\theta(\mathbf{0},t)-\theta(\mathbf{0},0))^2 \rangle}\times \nonumber \\
    && \left( e^{\frac 1 2 [\theta(\mathbf{0},t),\theta(\mathbf{0},0)]} -e^{-\frac 1 2 [\theta(\mathbf{0},t),\theta(\mathbf{0},0)]}\right),
\end{eqnarray}
since the effective action is Gaussian. In three dimensions, we have:
\begin{equation}\label{eq:theta2-3d}
  \langle \theta(\mathbf{0},0)^2 \rangle = \int \frac{\mathrm{d}^3
    \mathbf{q}}{(2\pi)^3} \frac{\pi u }{2K \omega(\mathbf{q})}
  \coth \left(\frac {\omega(\mathbf{q})}{2 T}\right),
\end{equation}
where:
\begin{equation}
    \label{eq:scha-dispersion}
    \omega^2(\mathbf{q}) = (uq_a)^2 +\frac{\pi u \kappa(T)}K\Bigl[2-\cos(\mathbf{q}\cdot\mathbf{b}) - \cos(\mathbf{q}\cdot\mathbf{c})\Bigr]
\end{equation}
The integral in Eq.~\eqref{eq:theta2-3d} is convergent,
yielding a non-zero expectation value for the order parameter of the AF ordered state $\sqrt{A_\perp} e^{-\langle \theta^2 \rangle/2}$. The response function~\eqref{eq:chiloc-scha} thus factorizes as:
\begin{eqnarray}
    \label{eq:scha-factorized}
    \chi^{\pm}_{\mathrm{loc.}}(t,T) = A_\perp e^{-\langle \theta^2 \rangle} \chi^{\pm}_{\mathrm{loc.}}(t,T=0) \Phi(t,T),
\end{eqnarray}
where:
\begin{eqnarray}\label{eq:chiloc-3d-gs}
    &&\chi^{\pm}_{\mathrm{loc.}}(t,T=0)= 2\Theta(t)
    \exp\Biggl\{-\frac{\pi}{4K} \int \frac{\mathrm{d}^2 \mathbf{q}_\perp}{(2\pi)^2}\nonumber\\
    &&\qquad\qquad\qquad\qquad\Bigl[Y_0\left[\omega_\perp(\mathbf{q}_\perp)t\right] -i J_0\left[\omega_\perp(\mathbf{q}_\perp)t\right] \Biggr\}\nonumber \\
    &&\qquad\qquad\qquad\times  \sin \left[\frac{\pi}{4K}\int  \frac{\mathrm{d}^2 \mathbf{q}_\perp
      }{(2\pi)^2}  J_0\left[\omega_\perp(\mathbf{q}_\perp)t\right]
  \right],
\end{eqnarray}
where
\begin{equation}\label{eq:omegaperp-def}
    \omega_\perp^2(\mathbf{q}) = \frac{\pi u \kappa(T)}K \Bigl[2-\cos(\mathbf{q}\cdot\mathbf{b}) - \cos(\mathbf{q}\cdot\mathbf{c})\Bigr],
\end{equation}
$Y_0$ and $J_0$ are Bessel functions~\cite{abramowitz1972} and
\begin{eqnarray}
    \label{eq:chiloc-3d-therm}
    &&\Phi(t,T)=\exp\Biggl[-\int \frac{\mathrm{d}^3\mathbf{q}}{(2\pi)^3} \frac{\pi u}{K\omega(\mathbf{q})}\nonumber\\
    &&\qquad\qquad\qquad\qquad\qquad\qquad\times\frac{1 - \cos\left[\omega(\mathbf{q}) t\right]}{e^{\omega(\mathbf{q})/T}-1}\Biggr]
\end{eqnarray}
By approximating $\omega_\perp(\mathbf{q}_\perp) \simeq v_\mathrm{sw}^{b,c} |\mathbf{q}_\perp|$, where $v_\mathrm{sw}^{b,c}= \sqrt{\pi u\kappa(T)/2K}=\pi u \sqrt{2\gamma(T)}$ and expanding for long times Eq.~\eqref{eq:chiloc-3d-gs}, before taking the Fourier transform, we obtain a tabulated integral,
Eq.~(11.4.35) in Ref.~\onlinecite{abramowitz1972}. Using the fluctuation dissipation theorem, and taking $\omega \to 0$, we recover~\eqref{eq:sw-nmr-t1}. With the same approximation for $\omega_\perp$, we obtain:
\begin{widetext}
    \begin{eqnarray}
    &&\chi^\pm_{\mathrm{loc.}}(t)=2 \Theta(t) \mathrm{e}^{-\frac{\pi}{12K} \left(\frac{T}{v_\mathrm{sw}^{b,c}}\right)^2} \exp\Biggl\{-\frac\pi {4K}    \Biggl[\frac{1}{2v_\mathrm{sw}^{b,c} t} \Bigl(Y_1(\pi v_\mathrm{sw}^{b,c} t) -i J_1(\pi v_\mathrm{sw}^{b,c} t)\Bigr) +\frac{ T^2}{(v_\mathrm{sw}^{b,c})^2 \sinh^2(\pi T t)}\Biggr]\Biggr\}\nonumber\\
    &&\qquad\qquad\qquad\qquad\qquad\qquad\qquad\qquad\qquad\qquad\qquad\qquad\qquad\qquad\qquad\qquad\qquad\times\sin \Bigl[\pi {8K v_\mathrm{sw}^{b,c} t} J_1(\pi v_\mathrm{sw}^{b,c} t) \Bigr].
    \end{eqnarray}
\end{widetext}
Expanding for $t\rightarrow +\infty$ and taking the Fourier transform leads to
\begin{eqnarray}
    \frac{1}{T_1^\perp} = \Biggl[\frac{m^\mathrm{AF}(T)}{v_\mathrm{sw}^{b,c}}\Biggr]^2 \frac{T}{4K} e^{-\frac\pi{12K}\left(\frac{T}{v_\mathrm{sw}^{b,c}}\right)^2},
\end{eqnarray}
so that:
\begin{equation}
    \label{eq:scha-t1-gs}
    \lim_{T\to 0} \frac 1 {T_1^\perp T} = \frac{1}{8\pi u J_\perp} e^{\frac{\mathcal{C}_4-\mathcal{C}_3}{8K}},
\end{equation}
and:
\begin{eqnarray}
    \label{eq:scha-t1-tpos}
    &&\frac 1 {T_1^\perp} = \frac{T e^{\frac{\mathcal{C}_4-\mathcal{C}_3}{8K}}}{8\pi u J_\perp} \left(\frac{-\frac 1 e \left(\frac T {T_c^\mathrm{SCHA}}\right)^4}{W_p\left[-\frac 1 e \left(\frac T {T_c^\mathrm{SCHA}}\right)^4\right]}\right.\nonumber\\
    &&\qquad\quad\quad\left.\times e^{\frac{8}{3\pi^3} \sqrt{-W_p\left[-\frac 1 e \left(\frac T {T_c^\mathrm{SCHA}}\right)^4\right]}} \right)^{\frac{1-4K}{8K}}.
\end{eqnarray}
Eq.~\eqref{eq:scha-t1-tpos} predicts a highly universal scaling, in which $1/(T_1^\perp T)$ depends on a universal function of $T/T_c$ raised to a power $1/(8K)-1/2$. Such a result is probably too universal, and reflects the limitations of the SCHA. It suggests however a plausible scaling law $1/T_1^\perp \sim T f(T/T_c,K)$. Plotting the expression~\eqref{eq:scha-t1-tpos} gives a nearly linear behavior of $1/T_1$ with temperature until $T\sim 0.6 T_c$. Past that point, different behaviors as a function of $K$ can be observed, with a modest enhancement of $1/T_1$ near the transition temperature. Clearly, the SCHA overestimates the order in the system when the temperature is getting close to $T_c$. This is a result of replacing the cosine potential with a quadratic potential. Such an approximation makes the gap $\Delta_s$ to create a soliton infinite. At low temperatures, since the soliton density goes as $\sim e^{-\Delta_s/T}$ this is a reasonable approximation. But near the critical temperature, the SCHA fails to account for the finite density of solitons and the resulting increase of phase space for relaxation. The zero temperature result~\eqref{eq:scha-t1-gs} gives $1/T_1^\perp\sim T/(J J_\perp)$ at low temperatures. For very weak interchain coupling, the spin wave velocity in the transverse direction is small, giving a large density of states at low energy. The reduction of the order parameter by the weak interchain coupling is insufficient to compensate this increase of density of states, explaining the increase of $1/T_1^\perp$ as $T\to 0$.
%

\end{document}